\documentclass{jfm}
\paperheight\pdfpageheight % Needed to tell hyperref the page size.
\paperwidth\pdfpagewidth   % Needed to tell hyperref the page size.
\usepackage{hyperref}
\usepackage{graphicx}
\usepackage[dvipsnames]{xcolor}
\usepackage{amsmath}
\usepackage[utf8]{inputenc}
\usepackage[normalem]{ulem} % For \sout
\usepackage{natbib}
\newcommand{\vect}[1]{\boldsymbol{#1}} % Vectors
\newcommand{\mat}[1]{\mathsfbi{#1}}    % Matrices
\newcommand{\eps}{\varepsilon}         % Small perturbation
\newcommand{\intd}{\mathrm{d}}         % straight d for derivatives and integrals
\newcommand{\I}{\mathrm{i}}            % \sqrt{-1}
\newcommand{\nondim}[1]{\tilde{#1}}    % Nondimensionalized variables
\newcommand{\nondimw}[1]{\widetilde{#1}} % Wide version of nondim

\newcommand{\hinfty}{h_\infty}

\shorttitle{Trapped Free Surface Waves for a Lamb--Oseen Vortex Flow}
\title{Trapped Free Surface Waves for a Lamb--Oseen Vortex Flow}

\author{E.~Zuccoli\aff{1},
E.~J.~Brambley\aff{1}\aff{2}\corresp{\email{E.J.Brambley@warwick.ac.uk}}
\and D.~Barkley\aff{1}}

\affiliation{\aff{1}Mathematics Institute, University of Warwick, UK
\aff{2}WMG, University of Warwick, UK}

\begin{document}

\maketitle

\begin{abstract}
Trapped surface waves have been observed in a swimming pool trapped by, and rotating around, the cores of vortices. To investigate this effect, we have numerically studied the free-surface response of a Lamb--Oseen vortex to small perturbations.  The fluid has finite depth but is laterally unbounded. The numerical method used is spectrally accurate, and uses a novel non-reflecting buffer region to simulate a laterally unbounded fluid.  While a variety of linear waves can arise in this flow, we focus here on surface gravity waves. We investigate the linear modes of the vortex as a function of the perturbation azimuthal mode number and the vortex rotation rate.
We find that at low rotation rates, linear modes decay by radiating energy to the far field, while at higher rotation rates modes become nearly neutrally stable and trapped in the vicinity of the vortex. 
While trapped modes have previously been seen in shallow water surface waves due to small perturbations of a bathtub vortex,
the situation considered here is qualitatively different owing to the lack of an inward flow and the dispersive nature of non-shallow-water waves.
We also find that for slow vortex rotation rates, trapped waves propagate in the opposite direction to the vortex rotation, whereas, above a threshold rotation rate, waves co-rotate with the flow. 
\end{abstract}

\begin{keywords}
%Authors should not enter keywords on the manuscript, as these must be chosen by the author during the online submission process and will then be added during the typesetting process (see http://journals.cambridge.org/data/\linebreak[3]relatedlink/jfm-\linebreak[3]keywords.pdf for the full list)
Surface gravity waves; Waves in rotating fluids.
\end{keywords}

\section{Introduction}

% General introduction
Rotating flows are rather ubiquitous in nature.  Examples range from domestic flows such as the bathtub vortex, to geophysical flows and Jupiter's Great Red Spot, to astrophysical flows such as accretion disks.  Rotating flows support the propagation of many different type of waves.  These are often viewed as small perturbations to an equilibrium state, also known as a base flow, and the behaviour of such waves depends both on the particular velocity field of the base flow and on the geometry of the problem under consideration. One important class of such waves are interfacial surface waves. These are waves forming at the interface between two fluids with different physical properties, and they often are strongly localized close to the interface.

\begin{figure}
    \centering
    \includegraphics[width=\textwidth]{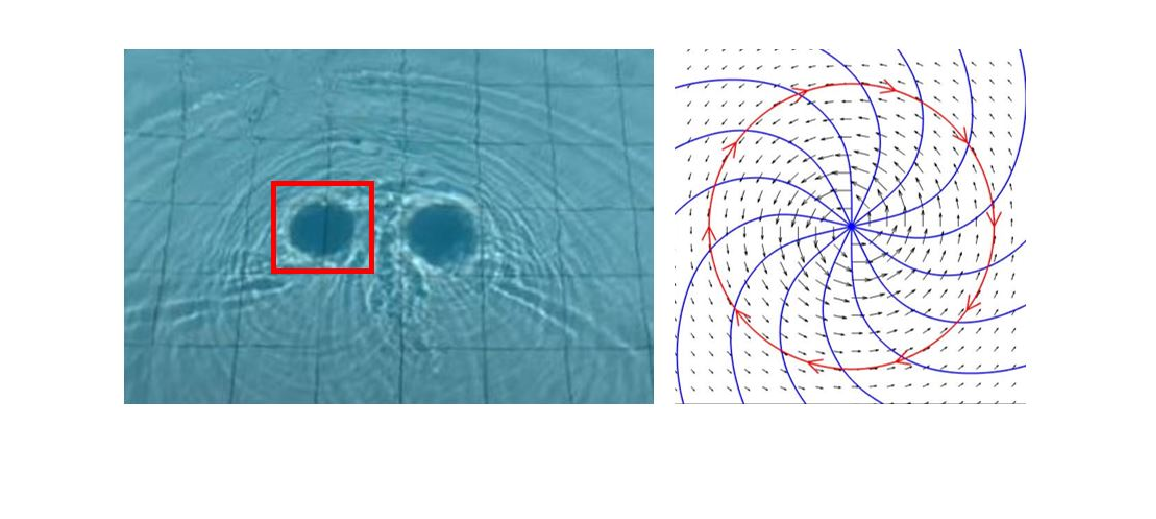}\vspace{-0.5in}
    \caption{The physical scenario motivating this work. Surface waves are trapped within a rotating vortex flow, but propagate in the direction opposite the direction of rotation.  Left: Photograph of two vortices in a swimming pool~\citep[a still taken from the video of][]{skipp-2020}.  Right: schematic of the flow and surface waves on the left-hand vortex.  Black arrows show the base velocity field. Blue lines show the wave crests. Red arrows show the direction of propagation of the waves.}
    \label{fig:pair_vortices_experiment}
\end{figure}

Our interest here is in free-surface waves on a vortex flow.
In fact, our initial motivation was out of curiosity concerning vortices and waves interacting in a swimming pool.  Changes in the surface height can be easily visualized by light and dark patterns on the swimming pool floor (for which the authors recommend a warm sunny climate and an outdoor swimming pool).  An ``experiment''~\citep{skipp-2020} created pairs of vortices by drawing a dinner plate through the water.  The initial waves generated by this disturbance disperse rapidly, leaving remarkably long-lived vortices with surface waves that appear trapped in the vortex, but which propagate around the vortex in the opposite direction to the vortex flow. A photograph of the phenomena, together with a schematic representation, is shown in figure~\ref{fig:pair_vortices_experiment};
interested readers are encouraged to follow the reference for a video.

The study of surface waves scattered by a vortex flow in a horizontally unbounded domain in a two-dimensional shallow-water setting with a bathtub vortex as a background flow was considered by \citet{patrick-2018} and \citet{patrick-2019}, in order to provide an analogy between fluid phenomena and black holes. Their study revealed the presence of modes that they called ``quasibound states'': normal modes having a very slow decaying rate in time and a spatial structure which remains trapped within the vortex core region,  similar to standing waves in a potential well. Such states are perhaps understandable for the bathtub vortex which includes an inflow towards the vortex.
The term ``quasibound'' references 
that temporally weakly-damped modes 
are not strictly bounded, and in fact
may have weak exponential growth at infinity in order to satisfy a causal radiation condition~(\citealp{briggs-1964,bers-1983}; \citealp[][pp.~17--18]{berti-2009}; \citealp{ledizes+billant-2009,riedinger+etal-2010}); despite this weak growth, such modes are
highly localized to a finite region (as depicted in figure~\ref{fig:free_surface_above_m7_varying_F} and discussed thereafter), and here we simply refer to these as \emph{trapped modes}.
Any slow decay of trapped waves may be due to dissipation (such as through viscosity), or by the wave not being perfectly trapped and ``leaking''  outgoing waves to infinity. Trapped waves could be excited either by an incoming propagating wave from infinity, or by any initial  disturbance (such as a dinner plate in the video depicted in figure~\ref{fig:pair_vortices_experiment}); here, we investigate the existence of trapped waves about a vortex, being agnostic as to the mechanism creating them in the first place.  We do not consider the related wave scattering problem.

We will study free surface waves on rotating fluid whose velocity $\vect{U_0}$ is that of a Lamb--Oseen vortex
\begin{equation}
\vect{U_0} = \vect{\hat{\theta}} \frac{\Gamma_0}{2\upi r} \Big(1-\exp\big\{{-r^2\!/a^2}\big\}\Big),
\label{eq:LO}
\end{equation}
where $r$ is the horizontal distance from the vortex centre, $\vect{\hat{\theta}}$ is a unit vector in the direction of rotation, and $\Gamma_0$ and $a$ are constants~\citep[see, e.g.][for details of the derivation of this flow]{drazin-2006}.   The Lamb--Oseen vortex is in fact an exact solution to the viscous Navier--Stokes equations if we take $a^2 = 4\nu t$, where $\nu$ is the kinematic viscosity, although for short timescales compared with the viscous diffusion timescale, $a$ may be approximated as constant.  Unlike the bathtub vortex, this base flow is not the result of vortex stretching by a downwards outflow.  
In that regard, it is more similar to a geophysical vortex, although here we do not consider a stratified fluid, nor do we assume the depth to be shallow.  
Unlike other vortices such as the Rankine vortex (made up of an inner solid body rotation and an outer potential vortex), which have been found to be unstable~\citep[e.g.][]{ford,mougel-2014},
the Lamb--Oseen vortex is usually considered to be a robust and stable coherent structure. 
\Citet{riedinger+etal-2010} showed that it can become unstable in a stratified environment, but with no stratification, \citet{fabre-2006} have shown that the Lamb--Oseen vortex is stable to axial-invariant perturbations. Their work highlighted the presence of critical layer waves, arising within a critical layer. However, because the critical layer provides a damping effect, they find the associated critical layer waves to be damped; we will not consider the critical layer further here.

% Previous literature

Free surface rotating waves have previously been studied computationally by \citet{mougel-2015,mougel-2017} in a confined geometry for background flows with particular forms, such as solid-body rotation or a potential vortex flow.  However, since a potential vortex has an infinite velocity at its centre, and a solid-body rotation has an infinite velocity at infinity, neither solid-body rotation nor a potential vortex are appropriate for the case we consider here.
Their studies reveal the presence of four main types of waves: surface gravity waves, inertial waves, Rossby waves, and centrifugal waves. The first three waves are observed for solid-body rotation, while centrifugal waves and gravity waves arise in a potential vortex flow.
Depending on the problem, each of these waves can be easily recognized and distinguished by its spatial structure and the associated frequency of oscillation.
Further examples of problems where surface gravity waves, inertial waves and Rossby waves appear can be found in \citet{johnson-1997}, \citet{greenspan-1969} and~\citet{mcwilliams-2006}, respectively. 
An interesting feature of the study on the potential vortex flow is the possibility of an instability of the base solution--- known as a Polygon Instability ---due to an interaction between the centrifugal and gravity waves.
This interaction mechanism was initially studied by \citet{tophoj+others} and subsequently by \citet{mougel-2017}.
The stability of vortex flows in bounded domains is however markedly different to the stability of vortex flows in unbounded domains, such as considered here, as the instability in bounded domains found by~\citet{tophoj+others} and~\citet{mougel-2017} involves the interaction of two types of wave modes, one of which is dependent on the outer boundary.  Instabilities of vortices on horizontally unbounded domains
have been found to occur within the shallow-water limit, as studied by \citet{ford}.  However, it is worth noting that the study of shallow-water rotating flows is rather different to the study of finite-depth rotating flows studied here, on two counts: firstly, shallow-water rotating flows can only rotate very slowly before the deformation in the free surface at the centre of the vortex touches the bottom boundary and a dry inner region is formed; and secondly, shallow-water waves are nondispersive and have a fixed wave speed, while finite-depth water waves are dispersive such that any wave speed is available to the system, and for example a wave can exist whose speed matches the flow speed at a given location.
Surface waves have been also studied by~\citet{huntNL-2015}, particularly considering the effects of an electric field on both linear and nonlinear inviscid, irrotational waves. Moreover, two-dimensional surface waves in electrohydrodynamics and magneohydrodynamics have been studied by~\citet{hunt-2015,hunt-2019,hunt-2021}, although in the present work we do not consider any electric or magnetic field, but instead we consider the full three-dimensional linear response to a prescribed swirling flow. 

% Discussion of non-reflecting boundary conditions
The present work considers a laterally unbounded domain (i.e.\ a domain that is in principle of infinite radial extent), with a finite depth, and thus assumes neither a shallow water nor a deep water limit.
In order to numerically study wave propagation problems in an unbounded domain, a non-reflecting boundary condition (NRBC) needs to be imposed at infinity. This is relatively straightforward in a two-dimensional shallow-water setting, since the dispersion relation for surface waves is known, and hence a characteristics analysis~\citep{john-1978} allows the imposition of an exact NRBC. Characteristic boundary conditions are, however, specific to classical hyperbolic wave equations~\citep{grote-1995,grote-2000,givoli-1989,bayliss-1980,majda-1977}. Such exact boundary conditions cannot be directly imposed on our three-dimensional problem, as our problem is not necessarily  hyperbolic, and it is not possible to find real characteristics. Moreover, surface waves on non-shallow water are dispersive, with the propagation speed of each mode depending on its frequency, which significantly complicates a NRBC even for hyperbolic problems~\citep[e.g.][]{lindquist-2012},  and can limit the range of wavenumbers being correctly treated~\citep[e.g.][]{wellens-2020}.
One possible solution to this problem is provided by absorbing layer methods. The idea behind this class of techniques is to surround the physical region with  computational layers through which waves are progressively damped such that they reach the edge of the computational domain with sufficiently small amplitude that any reflections back into the physical region will be negligible.
The absorbing layers need to vary gradually, however, as otherwise waves can spuriously reflect from the edges of the absorbing layers. Absorbing layer methods have been applied initially to waves problems in electromagnetics~\citep[e.g.][]{berenger-1994} and then extended to other physical fields like elasticity~\citealp[][]{hastings+schneider+broschat-1996, chew+liu-1996} and fluid dynamics~\citep{soderstrom+karlsson+museth-2010}. A shortcoming of absorbing layer formulations lies in the number of new unknowns introduced in the problem; this may be reduced using the formulation of~\citet{sim-2010} in the case of second order hyperbolic equations.
Here, a related but different technique is developed for the absorbing layer formulation in the far field, without introducing further unknowns in the mathematical problem.

% Plan of the paper
The paper is organised as follows: in section~\ref{sec:maths_model}, a general description is given of the mathematical model used to study perturbations to a Lamb--Oseen swirling flow, including a mathematical description of NRBCs. The resulting eigenvalue problem is then solved using a numerical procedure described in section~\ref{sec:num_method}.  The results of this numerical solution are described in section~\ref{sec:results}.  Finally, resulting conclusions and possible future directions for continued research are highlighted in section~\ref{sec:conclusions}.
% ============================================================
% END OF INTRODUCTION
% ============================================================
\section{Mathematical Model}\label{sec:maths_model}

Assuming that viscosity is negligible over the timescales of interest here, the governing equations are the incompressible Euler equations,
\begin{align}
\frac{\partial\vect{U}}{\partial t} + \vect{U\cdot \nabla}\vect{U} + \frac{1}{\rho}\bnabla P  + g\vect{\hat{z}} &= 0, &
\vect{\nabla \cdot U} &= 0,
\label{equ:euler}
\end{align}
where $\vect{U}$ is the fluid velocity, $P$ is the fluid pressure, $\vect{\hat{z}}$ is a unit vector in the vertical direction, and the constants $\rho$ and $g$ are the fluid density and the acceleration due to gravity respectively.  
The fluid is contained between a bottom boundary at $z=0$ and an upper free surface at $z = H$. The fluid must satisfy no penetration through the bottom boundary, giving $\vect{U\cdot\hat{z}} = 0$ at $z=0$.
Two boundary conditions, a kinematic and a dynamic boundary condition, must be satisfied along the free surface itself.  Here, we assume the fluid above the free surface to be dynamically passive, and in particular, to have a constant pressure $\bar{P}$.  Together, these give the boundary conditions
\begin{align}
\frac{\partial H}{\partial t} + \vect{U\cdot \nabla} H  &= \vect{U\cdot\hat{z}}
&&\text{and}&
P &= \bar{P}
&&\text{on}&
z &= H.
\label{equ:full_bcs}
\end{align}

We split the overall velocity and pressure into a steady purely swirling base flow and a small (magnitude $\eps$) time-dependent perturbation, where
\begin{align}
\vect{U} &= U_0(r)\vect{\hat{\theta}} + \eps(u_r\vect{\hat{r}} + u_\theta\vect{\hat{\theta}} + u_z\vect{\hat{z}}), &
P &= P_0(r,z) + \eps p, &
H &= h_0(r) + \eps h.
\end{align}

\subsection{The steady base flow solution}
\begin{figure}
    \centering
    \includegraphics[width=0.8\textwidth]{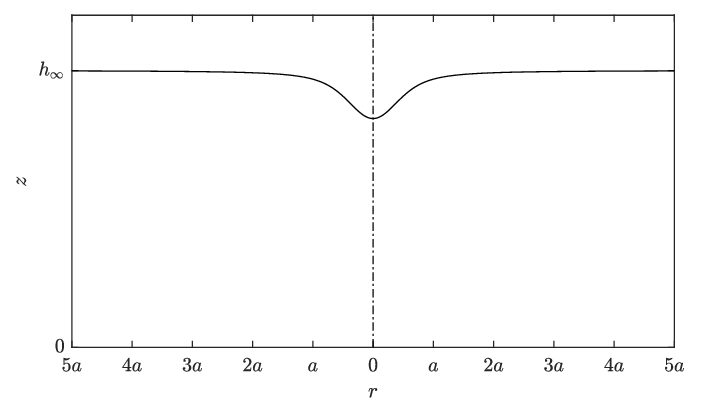}
    \caption{A typical section through a Lamb--Oseen vortex, showing the steady base flow free surface.}
    \label{fig:base_FS}
\end{figure}

A purely swirling steady base flow has a velocity given by $\vect{U_{0}} = U_{0}(r)\vect{\hat{\theta}}$. The governing equations and boundary conditions for this steady flow are satisfied provided we take
\begin{subequations}
\begin{align}
P_0(r, z) &= \bar{P} + \rho g \left( h_0(r) - z\right), \\
h_{0}(r) &= \hinfty - \frac{1}{g}\int_r^\infty \frac{U^{2}_{0}(r')}{r'}\,\intd r',
\end{align}%
\label{eq:P0_h0}%
\end{subequations}%
where $\hinfty$ is the depth of the fluid at $r=\infty$. This holds for any velocity profile $U_0(r)$. For the specific case of the Lamb--Oseen vortex considered here, we have 
\begin{equation}\label{lamb_oseen}
U_{0}(r) = \frac{\Gamma_{0}}{2\upi r}\Big(1-\exp\left(-r^2\!/a^2\right)\Big),
\end{equation}
where $a$ sets the radial size of the core and $\Gamma_{0}$ sets the circulation of the vortex.  Note that, for the Lamb--Oseen vortex, for small $r$, we have $U_0(r) \approx r\Gamma_0/2\upi a^2$, so that~\eqref{lamb_oseen} is a solid body rotation near the centre of the vortex, while for large $r$ we have $U_0(r)\approx \Gamma_0/2\upi r$, so that~\eqref{lamb_oseen} is a potential swirl far from the centre of the vortex.
A typical steady base flow free surface for the Lamb--Oseen vortex is shown in figure~\ref{fig:base_FS}.

\subsection{Perturbation Dynamics}
Waves arise when a small perturbation is introduced to the steady base solution.  By linearizing the governing equations~\eqref{equ:euler} about the base solution $(U_{0}, P_{0}, h_{0})$ given above, the governing equations for the perturbation are
\begin{subequations}\label{linear_euler_equations}
\begin{align}
D_{t}u_{r} - 2\Omega_{0}(r)u_\theta + \frac{1}{\rho}\frac{\partial p}{\partial r} &= 0, \\ 
D_{t}u_{\theta} + \frac{1}{r}\big(rU_0(r)\big)'u_{r} + \frac{1}{\rho r}\frac{\partial p}{\partial \theta} &= 0, \\
D_{t}u_{z} + \frac{1}{\rho}\frac{\partial p}{\partial z} &= 0, \\
\frac{1}{r}\frac{\partial}{\partial r}\left(ru_{r}\right) + \frac{1}{r}\frac{\partial u_{\theta}}{\partial \theta} + \frac{\partial u_{z}}{\partial z} &= 0,
\end{align}\end{subequations}
where $D_{t} = \partial_{t} + \Omega_{0}(r)\partial_{\theta}$ is the convective derivative and $\Omega_{0}(r) = U_{0}(r)/r$ is the steady base flow angular velocity.
(Here and throughout, primes denotes derivatives with respect to the argument for a function of only one variable).  Linearizing the boundary conditions~\eqref{equ:full_bcs} about the steady base flow leads to
\begin{subequations}\begin{gather}\begin{aligned}
u_{z} &= \frac{1}{\rho g}D_{t}p + h'_{0}u_{r}
&&\quad\text{and}\quad&
h &= \frac{p}{\rho g}
&&\quad\text{on}&
z &= h_{0}(r),
\end{aligned}\\
u_{z} = 0 \quad \text{on} \quad z = 0.
\end{gather}
\end{subequations}

Since our focus is on trapped waves and their formation,
in addition to these boundary conditions there is another implicit condition, which is that there are no waves entering the domain from $r=\infty$, and thus only outgoing waves are allowed at $r=\infty$. 
This rather subtle condition will become more concrete when we truncate the domain to finite $r$ in order to numerically solve the above equations.

Before we progress further, we now make two changes to the governing equations for the perturbation.  Firstly, since the governing equations for the perturbation are linear, we may assume a modal solution of the form
\begin{subequations}\label{equ:modal}\begin{align}
\vect{u} &= \left[u(r, z)\vect{\hat{r}} + v(r, z)\vect{\hat{\theta}} + w(r, z)\vect{\hat{z}}\right]\exp(-\I\omega t + \I m\theta),
\\
p &= \phi(r, z)\exp(-\I\omega t + \I m\theta),
\end{align}\end{subequations}
where $m$ is restricted to integer values, since the solution must be $2\upi$ periodic, while $\omega$ in general will be complex, with $\Real(\omega)/2\upi$ being the oscillation frequency and $-\Imag(\omega)$ being the decay rate.  The eigenvalue problem that will eventually result will have $\omega$ as the eigenvalue to be found.

Secondly, we re-write the governing equations and boundary conditions in a non-dimensional form.  To do so, we choose the reference lengthscale to be $a$, the scale of the vortex core (equation~\eqref{eq:LO} and figure~\ref{fig:base_FS}). The reference time scale is that given by this lengthscale and gravity: $\sqrt{a/g}$.  
The velocity scale is thus $\sqrt{ag}$.
All dimensional variable may then be expressed in terms of a nondimensional variable (denoted by a tilde), as
\begin{equation}\begin{gathered}
(r,\theta,z) = (a\nondim{r}, \nondim{\theta}, a\nondim{z}), \qquad
t = \sqrt{\frac{a}{g}}\nondim{t}, \qquad
U_{0}(r) = \frac{\Gamma_{0}}{2\upi a}\nondimw{U}_{0}(\nondim{r}), \qquad
\Omega_0 = \frac{\Gamma_{0}}{2\upi a^2}\nondim{\Omega_0},
\\
h = a\nondim{h}, \qquad
p = \rho ag\nondim{p}, \qquad
\omega = \sqrt\frac{g}{a}\nondim{\omega}, \qquad
\vect{u} = \sqrt{ag}\vect{\nondim{u}}.
\end{gathered}\end{equation}

There are two physical parameters that are not scaled to unity by this nondimensionalization: these may be thought of as the strength of the vortex
and the depth of the fluid at infinity,
given respectively as
\begin{align}
F &= 
\frac{\Gamma_0}{2\upi\sqrt{ga^3}}, &
\nondim{h}_\infty &= \frac{\hinfty}{a}.
\label{froude_number}
\end{align}
$F$ is the Froude number and sets the nondimensionalized velocity of the vortex. Hence, $F \to 0$ corresponds to a slow vortex with negligible steady surface height variation, while $F \to \infty$ corresponds to a fast vortex with significant steady surface height variation, as can be seen from~(\ref{dimensioness_base_free_surface}\textit{c}) below. The dimensionless depth $\nondim{h}_\infty$ is exactly that, so that the limit $\nondim{h}_\infty \to 0$ corresponds to the shallow-water limit and $\nondim{h}_\infty \to \infty$ corresponds to the deep-water limit.  Care is needed, however, in considering the shallow-water limit: in what follows, we will assume that the steady fluid height never reaches zero, so that the bottom stays wetted, and consequently the shallow water limit $\nondim{h}_\infty \to 0$ must be taken together with the slow wide vortex limit $F\to 0$ such that $\nondim{h}_\infty/F^2$ is bounded away from zero, as can also be seen from~\eqref{dimensioness_base_free_surface} below.
The azimuthal wavenumber $m$ is also a dimensionless parameter, and represents the rotational symmetry of the solution being investigated.

Dropping the tildes, the complete nondimensional eigenvalue problem is
\begin{subequations}\label{equ:linearized}\begin{align}
\big({-\I}\omega + \I m F \Omega_{0}(r)\big)u - 2F\Omega_{0}(r)v + \frac{\partial \phi}{\partial r} &= 0, \\
\big({-\I}\omega + \I m F \Omega_{0}(r)\big)v + \frac{F}{r}\big(rU_0(r)\big)'u + \frac{\I m}{r}\phi &= 0, \\
\big({-\I}\omega + \I m F \Omega_{0}(r)\big)w + \frac{\partial \phi}{\partial z} &= 0, \\
\frac{1}{r}\frac{\partial}{\partial r}\left(ru\right) + \frac{\I m}{r}v + \frac{\partial w}{\partial z} &= 0,
\label{equ:linearized-continuity}
\end{align}\end{subequations}
together with the boundary conditions of no incoming modes at $r=\infty$, and
\begin{subequations}\label{equ:linearized-bcs}\begin{align}
w &= 0, &\text{on} \quad z &= 0,\\
w &= \big({-\I}\omega + \I m F \Omega_{0}(r)\big)\phi + F^{2} r\Omega^{2}_{0}(r)u,
&\text{on} \quad z &= h_{0}(r).
\end{align}\end{subequations}
For the Lamb--Oseen vortex, in dimensionless terms
\refstepcounter{equation}\[
U_0(r) = \frac{1 - \exp(-r^2)}{r}, \quad
\Omega_0(r) = \frac{U_0(r)}{r}, \quad
h_{0}(r) =  h_{\infty} - F^2 \int_r^\infty \frac{U^{2}_{0}(r')}{r'}\,\intd r'.
 \eqno{(\theequation{\mathit{a},\mathit{b},\mathit{c}})}\label{dimensioness_base_free_surface}
\]

One consequence of the harmonic assumption~\eqref{equ:modal} is the creation of a so-called ``critical layer'' at a radial location $r=r_c$, given by $D_t = -\I\omega + \I mF\Omega_0(r_c) = 0$.  It is not immediately obvious from~\eqref{equ:linearized} that anything particularly special occurs at the critical layer, as in no equation does it cause the highest derivative to vanish, and therefore our numerics described in~\S\ref{sec:num_method} has no difficulty in this case.  However, in fact the critical layer is related to behaviour that is not of the harmonic form assumed in~\eqref{equ:modal}; investigating the effect of the critical layer requires a different mathematical and numerical technique~\citep[such as Frobenious series; see, by way of example,][]{king+brambley-2022}, which is beyond the scope of this paper.  However, we comment in passing that \citet{fabre-2006} found that disturbances related to the critical layer are necessarily damped, suggesting that the undamped trapped wave modes we are interested in here are not related to the critical layer.

\section{Numerical methods}\label{sec:num_method}

\subsection{Absorbing layer for 3D incompressible Euler equations}
\label{section:absorbing-layer}

In practice, we solve the eigenvalue problem in a finite computational domain and hence introduce an artificial boundary at finite radius $R$. The far-field boundary condition of no incoming waves becomes a non-reflecting boundary condition (NRBC) at this boundary. However, exact NRBCs are generally derived using the method of characteristics, but this is not available to us since~\eqref{equ:linearized-continuity} is not hyperbolic.  A widely used alternative is based on damping layers or even perfectly matched layers. The idea is to introduce a ``damping layer'' --- also called an ``absorbing layer'' --- surrounding the physical outer boundary. In the damping layer, waves are progressively damped until their amplitude is sufficiently small that reflections from the outer boundary do not reenter the physical domain. Outgoing waves are effectively absorbed in the layer. The boundary condition at the outer boundary of the computational domain is then unimportant, and is typically taken to be either a Dirichlet-type or Neumann-type boundary condition.

\begin{figure}%
\centering%
    \includegraphics[height=0.267\textheight]{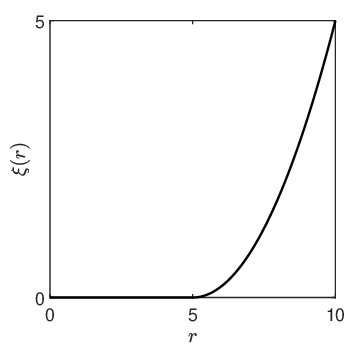}%
    \includegraphics[]{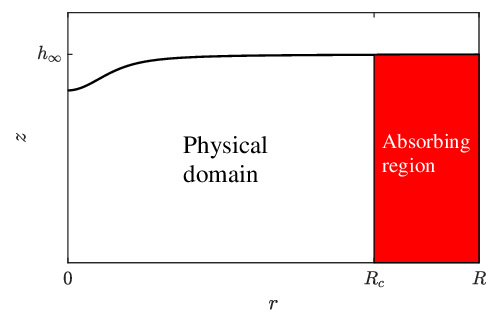}%
\caption{Left: profile of the quadratic damping function defined in~\eqref{equ:definition_damping_function} with parameters: $\bar{\xi} = 5$, $R_c = 5$ and $R = 10$. Right: schematic of the computational domain in the $r-z$ plane.}%
\label{fig:damping_region_and_function}%
\end{figure}

To add damping to our governing equations, we add a ``damped compressibility'' term into the continuity equation~\eqref{equ:linearized-continuity}, which then becomes
\begin{equation}
\xi(r)\phi + \frac{1}{r}\frac{\partial}{\partial r}\left(ru\right) + \frac{\I m}{r}v + \frac{\partial w}{\partial z} = 0,
\label{equ:modified-continuity}
\end{equation}
where $\xi(r)$ is a sufficiently smooth function that is identically zero outside the damping region.  This is motivated by the introduction of damping in the acoustic wave equation following \citet[][pp. 81-82]{gao+others}, and is explained in further detail in appendix~\ref{appendix:damping_layer_form}.  In what follows, we find good results using the simple form for $\xi(r)$ given by
\begin{equation}
\xi(r) = 
\begin{cases}
0, & r < R_{c}, \\
\bar{\xi}\left(\dfrac{r - R_{c}}{R - R_{c}}\right)^{2}, & R_{c} \le r \le R.
\end{cases}
\label{equ:definition_damping_function}
\end{equation}
This gives a physical region, $r < R_{c}$, with no damping, and computational region $r\in[R_c,R]$ with a damping strength governed by the constant $\bar{\xi}$. We then impose a Dirichlet boundary condition on the pressure at an artificial boundary $r = R \gg 1$. We show in section~\ref{sec:causality} that this leads to the correct causal behaviour.
A representation of the damping function as well as the subdivision of the numerical domain is shown in figure~\ref{fig:damping_region_and_function}. 

The governing equations~\mbox{(\ref{equ:linearized}\textit{a--c})} together with the modified continuity equation~\eqref{equ:modified-continuity} and boundary conditions~\eqref{equ:linearized-bcs} form an eigenvalue problem to solve for the allowable frequencies $\omega$ permitting a nonzero modal solution.

\subsection{\texorpdfstring{Numerical discretization}{Numerical discretization}}

In order to solve~(\ref{equ:linearized}\textit{a--c},\ref{equ:linearized-bcs},\ref{equ:modified-continuity}), we used a Galerkin spectral method and expand with a combination of Legendre polynomials as basis functions in both the radial and axial coordinate in order to satisfy the Dirichlet boundary conditions. We also remap the domain from the domain $D = [0, R]\times[0, h_{0}(r)]$ to the computational square domain $S = [-1, 1]\times[-1, 1]$ to account for the shape of the computational domain with a variable surface height $h_0(r)$. We then obtain the weak formulation of the problem.  Full details are given in appendix~\ref{appendix:numerics}.  The discretized problem ends up being of the form
\begin{equation}
\begin{aligned} &
\mat{A}\vect{w} = \omega \mat{B}\vect{w}, 
\end{aligned}
\label{equ:discrete_evals_problem_fist_written}
\end{equation}
with $\vect{w} = (u_{ij}, v_{ij}, w_{ij}, \phi_{ij})$ representing our array containing the spectral coefficients of each unknown, $\mat{A}$ and $\mat{B}$ being matrices of order $4N_{x}N_{y} \times 4N_{x}N_{y}$, where $N_x$ and $N_y$ are the number of Legendre polynomials used in the radial and vertical directions respectively. This discretized problem may then be solved using any numerical eigenvalue solver; here, we use the $\texttt{eig}$ solver in \textsc{Matlab}.
We sort eigenvalues by imaginary part and focus on the leading eigenvalues, i.e.\ those with most positive imaginary part. These correspond to unstable modes, should there be any, or the least damped stable modes, if not.

Not all solutions to the discretized problem~\eqref{equ:discrete_evals_problem_fist_written} correspond to solutions to the continuous problem being approximated, however.  To remove under-resolved eigenmodes and spurious eigenvalues, the numerical solutions to~\eqref{equ:discrete_evals_problem_fist_written} are filtered, as described in the appendix~\ref{sec:resolvedness}. 
The numerical scheme has been validated against three published cases: two with walls at finite radius \cite{mougel-2014,mougel-2015}, and one unbounded but in shallow water \cite{ford}. We obtain unstable modes where they are known to exist. Details are given in appendix~\ref{appendix:code_validation_SB_rotation}.

\subsection{Spurious reflected modes}
\label{sec:spurious-reflection}

The finite numerical domain and damping region introduces another source of spurious modes besides those coming from the numerical discretization, namely those modes which are well-resolved but which include a significant reflection from either the damping boundary at $r=R_c$ or the truncation boundary at $r=R$.  Such spurious eigenmodes are affected by the values of $R_c$ and $R$, and by the amount of damping $\bar{\xi}$, whereas good approximations to the modes on the infinite domain should be insensitive to these values.  Since -$\Imag(\omega)$ is the decay rate of the mode, variations in damping typically have a strong effect on $\Imag(\omega)$ for spurious reflected modes.  We may therefore remove these spurious reflected modes by running our numerical code twice: the first time with a suitably chosen amount of damping $\bar{\xi}$, and the second time with twice that amount of damping $2\bar{\xi}$.  Only those modes whose eigenvalues do not change significantly with the change in the damping coefficient are retained (as measured using the same metric~\eqref{resolv_cond_eigenvalues} used for the numerical resolution).

\subsection{Numerical convergence study}
\label{sec:convergence}

\begin{figure}
  {\centering\includegraphics[width=0.49\textwidth]{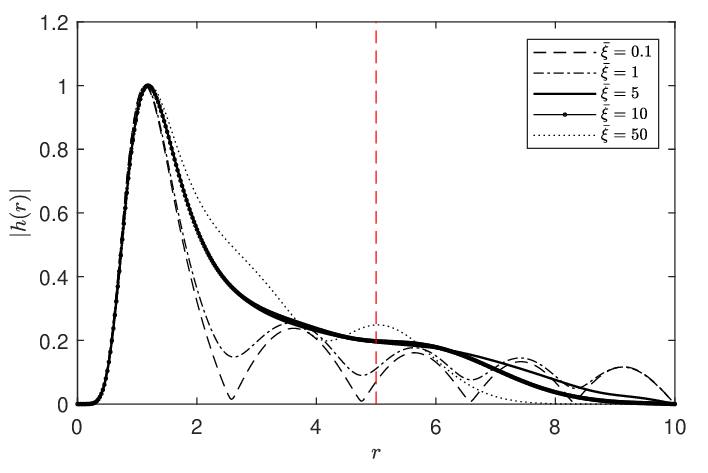}%
  \includegraphics[width=0.49\textwidth]{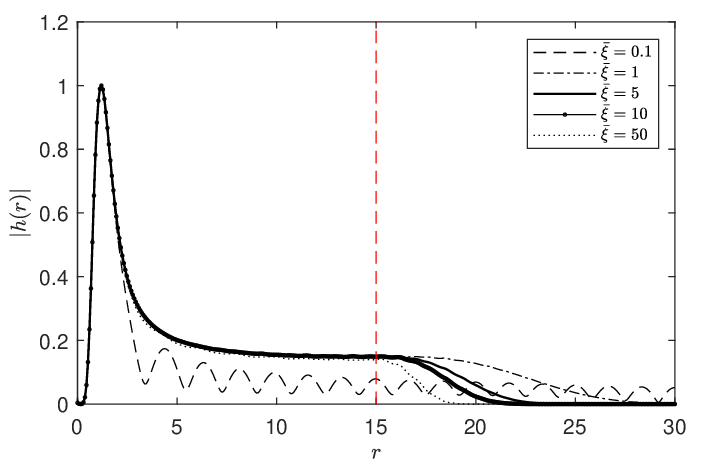}}
  \caption{Eigenfunctions for $m = 7$, $F = 0.3$, and different values of $\bar{\xi}$. A vertical red-dashed line indicates the radial point $R_c$ where the damping effect begins.  Left: a small computational domain $R_{c} = 5$, $R = 10$.  Right: a large computational domain $R_c=15$, $R=30$.}
\label{fig:conv_wrt_xibar_eigfunc}
\end{figure}

The first convergence study involves varying the amount of damping in the damping layer and checking that the eigenvalues do not change, nor does the shape of the corresponding eigenfunctions in the physical domain $r < R_{c}$.  Here we present the convergence results for $m = 7$ and $F = 0.3$, 
since for these parameters the system supports radiating (outwardly propagating) modes, and such modes provide the most stringent test of a non-reflecting boundary condition.
For varying magnitudes of damping $\bar{\xi}$, the radiating eigenfunction is plotted in figure~\ref{fig:conv_wrt_xibar_eigfunc},
and the corresponding eigenvalues are given in table~\ref{table:conv_wrt_xibar}.
\begin{table}
\centering
 \begin{tabular}{c c c c c c} 
 \hline
 $\bar{\xi}$ & 0.1 & 1 & 5 & 10 & 50 \\ [0.5ex]
 \hline\hline
 $\Real(\omega)$ & -1.3911 & -1.3894 & -1.3766 & -1.3769 & -1.3688 \\
 $\Imag(\omega)$ & $-8 \times 10^{-4}$ & -0.0085 & -0.0133 & -0.0138 & - 0.0123 \\
 \hline
 \end{tabular}
 \caption{Eigenvalues as function of the amount of damping for $m=7$ and $F = 0.3$, $R_{c} = 5$ and $R = 10$.}
 \label{table:conv_wrt_xibar}
\end{table}%
Two domain sizes are shown in figure~\ref{fig:conv_wrt_xibar_eigfunc}: $R_c=5$ and $R=10$ are the values used for the results in section~\S\ref{sec:results}; and $R_c=15$ and $R=30$ give an extended domain so that the effects of damping and resonance can be seen more clearly. The damping clearly influences the eigenfunction shape, and for $\bar{\xi}\in\{0.1,1\}$ a clear standing wave shape is seen in figure~\ref{fig:conv_wrt_xibar_eigfunc}. For $\bar{\xi}\in\{5,10\}$, the eigenfunctions are practically identical in the physical domain $0 < r < R_c$, and only differ in the damping layer $R_c\leq r \leq R$.  For $\bar{\xi}=50$, however, the damping is too strong, and little oscillations can be seen for $r<R_c$, suggesting wave reflection by the edge of the damping layer. This is also supported by the eigenvalues in table~\ref{table:conv_wrt_xibar}, which show the sensitivity of $\Imag(\omega)$ to variations in damping strength, as expected.

\begin{figure}
  {\centering\includegraphics[width=0.49\textwidth]{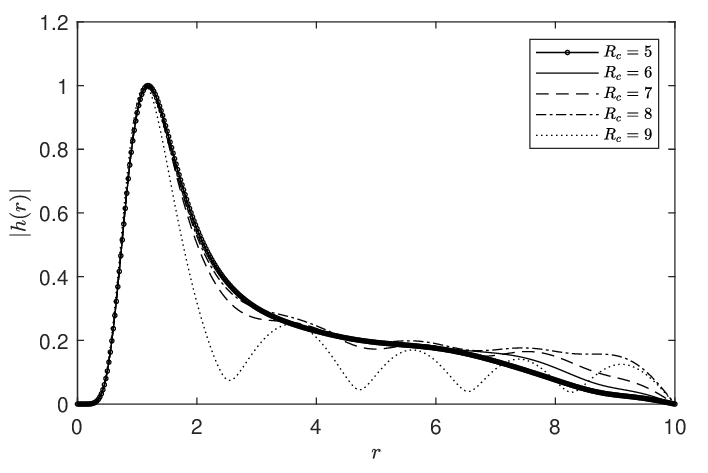}%
  \includegraphics[width=0.49\textwidth]{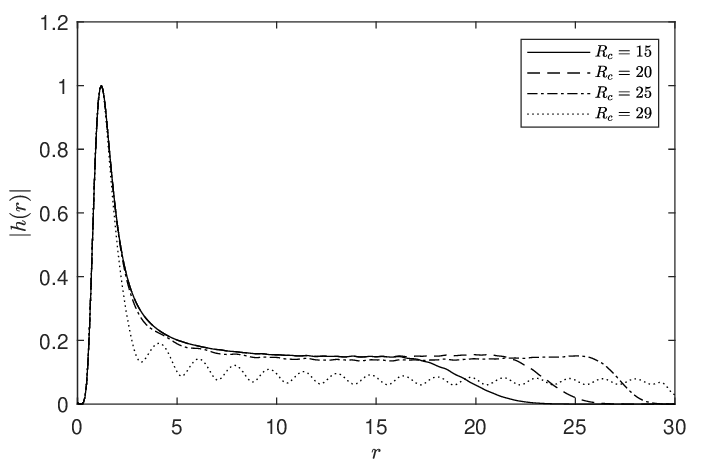}}
  \caption{Eigenfunctions for $m = 7$, $F = 0.3$, $\bar{\xi} = 5$, and different values of $R_{c}$, computed using a small domain $R=10$ (left) and a larger domain $R=30$ (right).}
\label{fig:conv_wrt_Rc_eigfunc}
\end{figure}

The second convergence study involves varying the width of the damping layer whilst maintaining the same size of the computational domain and a fixed damping strength $\bar{\xi} = 5$.  The eigenfunctions are displayed in figure~\ref{fig:conv_wrt_Rc_eigfunc},
and the corresponding eigenvalues in table~\ref{table:conv_wrt_Rc}.
\begin{table}
\centering
 \begin{tabular}{c c c c c c c c} 
 \hline
 $R_{c}$ & 5 & 6 & 7 & 8 & 9\\ [0.5ex]
 \hline\hline
 $\Real(\omega)$ & -1.3766 & -1.3767 & -1.3802 & -1.3788 & -1.3918 \\
 $\Imag(\omega)$ & -0.0133 & -0.0131 & -0.0133 & - 0.0143 & -0.0045 \\
 \hline
 \end{tabular}
 \caption{Eigenvalues as function of the initial position of the absorbing layer for $m=7$ and $F = 0.3$, $\bar{\xi} = 5$ and $R = 10$.}
 \label{table:conv_wrt_Rc}
\end{table}%
For most results in figure~\ref{fig:conv_wrt_Rc_eigfunc} the eigenfunctions can be seen to be very similar in the physical domain $0<r<R_c$ and to decay smoothly in $r$ in the damping layer, while for a damping layer of width $R-R_c=1$ a standing wave pattern can be seen for both domain sizes, implying significant wave reflection from the domain boundary at $r=R$. Again, this is also seen for the variations in the eigenvalue in table~\ref{table:conv_wrt_Rc}, with again $\Imag(\omega)$ being particularly sensitive.

\begin{figure}
  \centerline{\includegraphics[width=0.8\textwidth]{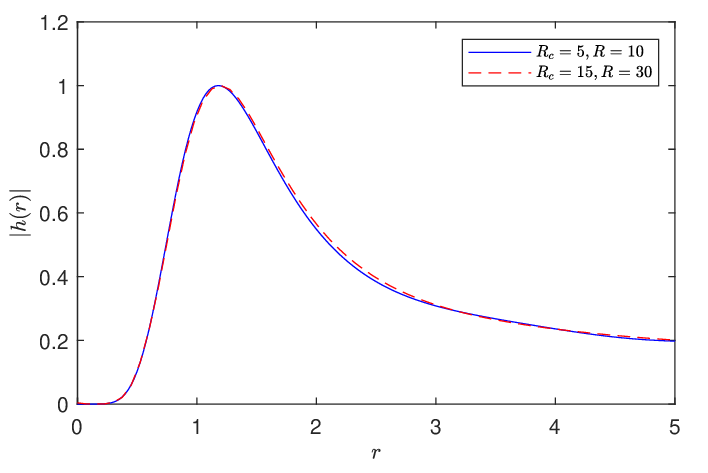}}
  \caption{Eigenfunction for $m = 7$, $F = 0.3$, $\bar{\xi} = 10$ computed using two different domain sizes: $R_c = 5$ and $R = 10$ for the blue solid curve; and $R_{c} = 15$, $R = 30$ for the red dashed curve.}
\label{fig:comp_modes_Rc5_Vs_Rc15}
\end{figure}

Finally, in figure \ref{fig:comp_modes_Rc5_Vs_Rc15}
we compare the radiating mode for $m = 7$, $F = 0.3$ computed using two different size of the domain; $R_c = 5$, $R = 10$ on one hand and $R_c = 15$, $R = 30$ on the other. The two modes match very well over the physical interval $r \in [0, 5]$. This good match demonstrates that the damping layer successfully allows our numerical simulation on a small interval to reproduce results that would have been obtained using a larger interval, thus allowing our numerics to emulate an infinite unbounded domain using a finite computational domain.

\subsection{Choice of numerical parameters}

Based on numerical convergence studies, for the results that follow we take $R = 10$ and $R_{c} = 5$, with $\bar{\xi} = 5$ (to which results are compared to $\bar{\xi}=10$).  This choice is motivated by the need for sufficiently high resolution to resolve all modes of interest in the range of Froude numbers and azimuthal wavenumbers considered.  The eigenvalue tolerance and eigenfunction resolvedness tolerance are taken to be $\mathrm{tol} = 10^{-2}$ and $10^{-1}$ respectively, whereas other numerical parameters are taken to be $N_{x} = 50$, $N_{y} = 20$, $b_{x} = 12$ and $b_{y} = 4$, as explained in more details in appendix \ref{appendix:numerics}.
For the majority of results presented here, we use a fluid depth of $h_\infty = 5$. This is the depth shown in figure~\ref{fig:base_FS}. This value is large enough to allow for a wide range of Froude numbers ($0 \leq F < \sqrt{h_{\infty}/\log(2)} \simeq 2.68$) without forming a dry region near $r=0$, while small enough to exhibit finite-depth effects.

% ======================================================
% END NUMERICAL METHOD
% ======================================================

\section{Results}
\label{sec:results}

We compute leading (i.e.\ least damped) eigenmodes for a range of Froude numbers and for a range of azimuthal wave numbers as large as $m=20$, and find that above $m = 6$ surface gravity waves dominate.  We shall first present the surface waves for a representative case, wavenumber $m=7$, and discuss in detail how the modes and eigenvalues depend on Froude number for this case. Following this, we consider the dependence of modes and eigenvalues on the azimuthal wave number. 

\subsection{Representative case \texorpdfstring{$m=7$}{m=7}} % texorpdfstring in order to avoid compilation warning (involving pdf list of bookmarks)

We begin with a detailed description of the case azimuthal wavenumber $m=7$. The reason for taking such value of $m$ is dictated both by the explanatory picture in \citet[pp.6]{patrick-2018}, who show a picture of the spiral structure of a normal mode solution for sufficiently high azimuthal wavenumber, as well as by the reasonable amount of computational time needed to get the leading surface waves eigenmodes. Indeed, we find numerically that the spatial structure of surface gravity waves becomes thinner and more localized close to the free surface as $m$ increases. Following particularly the latest argument, the case of $m=7$ has been taken as a reference study case and in the following we are going to show most of the interesting features applied to this case.

\begin{figure}%
\centering%
    \includegraphics[]{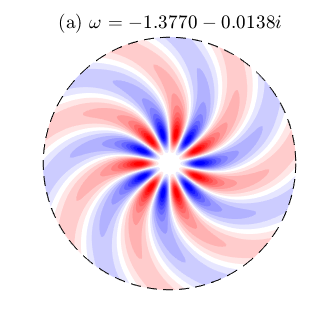}%
    \includegraphics[]{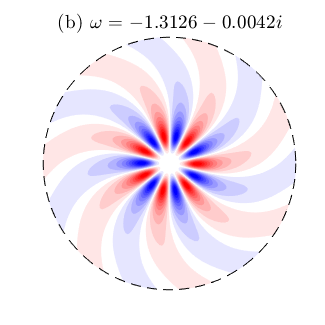}%
\par%
    \includegraphics[]{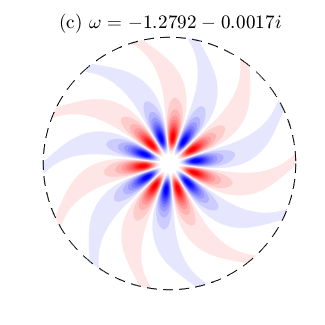}%
    \includegraphics[]{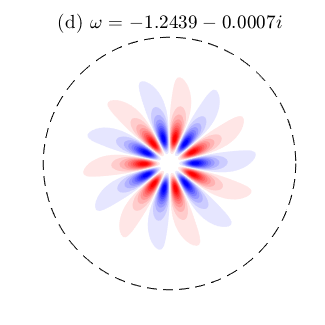}%
\par%
    \includegraphics[]{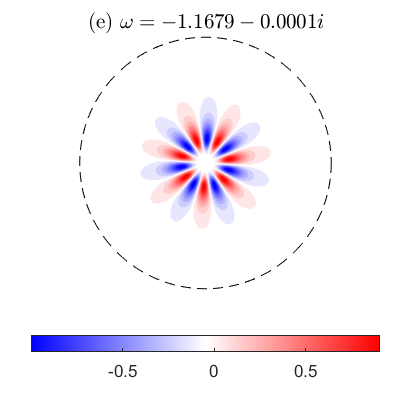}%
\caption{Plots of the free surface height $h(r, \theta, t=0) = \Real[\phi(r, h_{0}(r))\exp\{\I m\theta\}]$ for $m = 7$. 
(a) $F = 0.3$. (b) $F = 0.32$. (c) $F = 0.33$. (d) $F = 0.34$. (e) $F = 0.36$.
All modes displayed rotate clockwise, i.e.\ against the vortex flow.  Animations of~(a) and~(e) are given in Movies~1 and~2 in the online supplementary material.}%
\label{fig:free_surface_above_m7_varying_F}%
\end{figure}

\begin{figure}%
\centering%
    \includegraphics[]{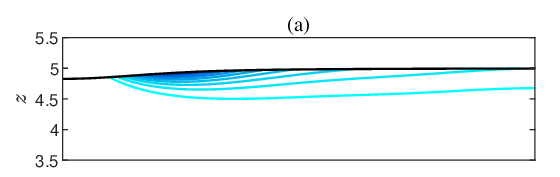}\par%
    \includegraphics[]{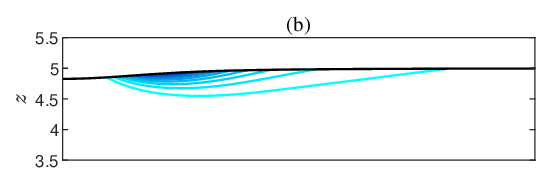}\par%
    \includegraphics[]{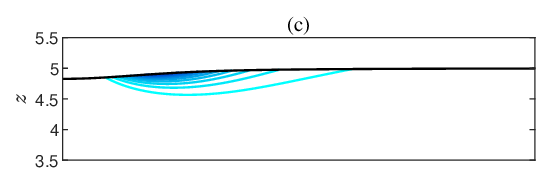}\par%
    \includegraphics[]{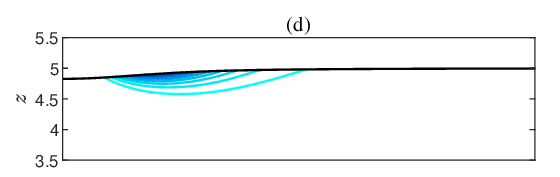}\par%
    \includegraphics[]{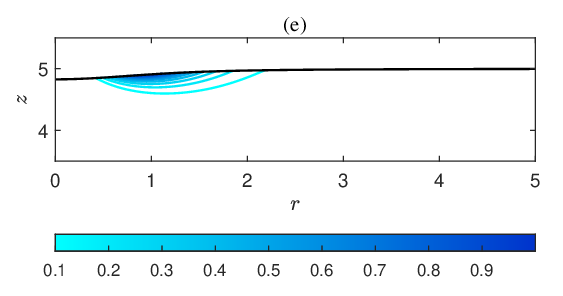}%
\caption{Plots of the pressure distribution $|\phi_{m}(r, z)|$ for $m = 7$. (a) $F = 0.3$. (b) $F = 0.32$. (c) $F = 0.33$. (d) $F = 0.34$. (e) $F = 0.36$.  Animations of~(a) and~(e) are given in Movies~1 and~2 in the online supplementary material.}%
\label{fig:pressure_contour_radiating_regime}%
\end{figure}

One of the key results of our study is the change in character of modes as function the rotation rate of the vortex. Specifically, we characterize two extremes of modes: radiating modes and trapped modes.
The spatial structure of the radiating modes has a characteristic spiral shape, as shown in figure~\ref{fig:free_surface_above_m7_varying_F}a,
% -----------------------------------------
% FREE SURFACE FROM ABOVE
% -----------------------------------------
% -----------------------------------------
and a long tail in the radial direction, as shown in figure~\ref{fig:pressure_contour_radiating_regime}a.
Such modes behave as travelling waves along the radial direction, and radiation to infinity is responsible for the dissipation of the initial energy put into the system to excite the mode.
(An animation of the radiating mode in figures~\ref{fig:free_surface_above_m7_varying_F}a and~\ref{fig:pressure_contour_radiating_regime}a is given as Movie~1 in the online supplementary material.)
In contrast, trapped modes are confined to a bounded region in the radial direction, and without the spiraling structure seen for the radiating modes. Those modes
behave as standing waves along the radial direction and dissipate extremely slowly in comparison.  Examples can be seen in figures~\ref{fig:free_surface_above_m7_varying_F}e and~\ref{fig:pressure_contour_radiating_regime}e.
(An animation of the the trapped mode in figures~\ref{fig:free_surface_above_m7_varying_F}e and~\ref{fig:pressure_contour_radiating_regime}e is given as Movie~2 in the online supplementary material.)
The distinction between radiating and trapped modes can further be seen in the eigenvalues. Indeed, radiating modes have eigenvalues with a significant negative imaginary part, indicating exponential decay in time, whereas trapped modes have a negligible imaginary part and so do not appreciably decay in time.

\begin{figure}%
    \centering%
    \includegraphics[width=0.8\textwidth]{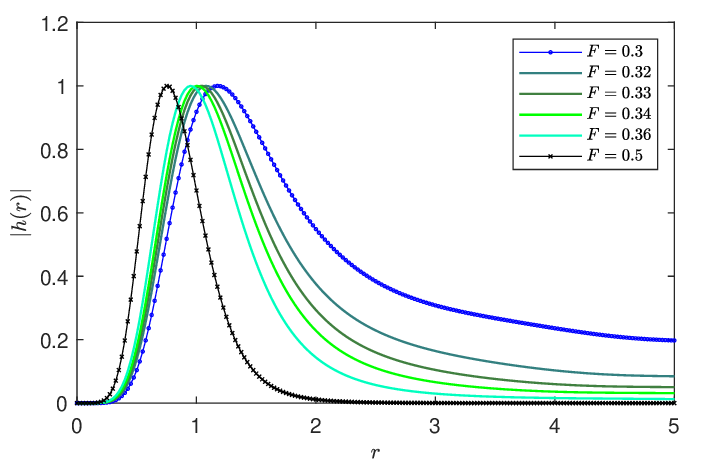}%
  \caption{Modulus of the free surface height along the radius for $m = 7$ in the transition regime $F \in [0.3, 0.36]$ from radiating to trapped modes.
  The trapped mode at $F = 0.5$ is also shown for comparison.
  }%
\label{fig:free_surface_m7_varying_F}%
\end{figure}

Figures \ref{fig:free_surface_above_m7_varying_F} and \ref{fig:pressure_contour_radiating_regime} show representative examples of the continuous, but rapid, transition from radiating modes to trapped modes as the Froude number (the dimensionless rotation rate of the vortex), increases. This transition between radiating and trapped modes is also seen in figure~\ref{fig:free_surface_m7_varying_F}
%
% Single plot of |h(r, 0)| for different F.  
%
where the modulus of the free surface height is plotted as a function of $r$ for different Froude numbers.

\begin{figure}%
\centering%
\includegraphics[]{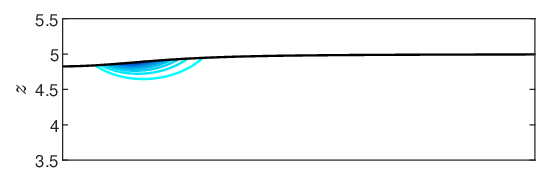}
\includegraphics[]{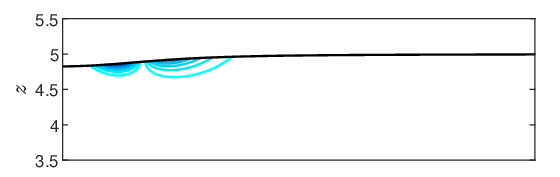}
\includegraphics[]{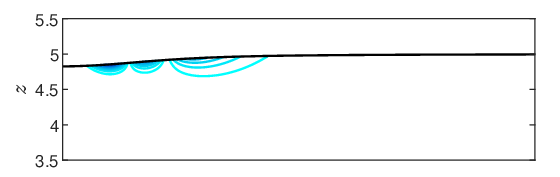}
\includegraphics[]{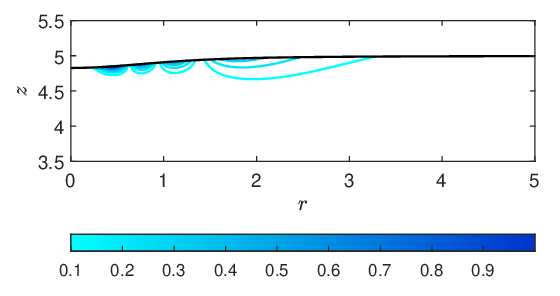}%
\par%
\caption{Modulus of the pressure $|\phi_{m}(r, z)|$ for $m = 7$ and $F = 0.5$. We show the four least stable eigenmodes for such parameters, having one, two, three and four peaks respectively.}%
\label{fig:peaks_eigenmodes}%
\end{figure}

\begin{figure}%
\centering%
(a)\raisebox{-0.9\height}{\includegraphics[width=0.8\textwidth]{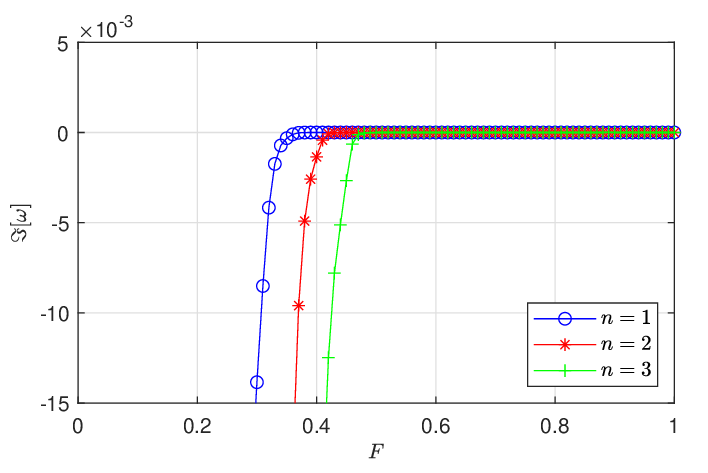}}\par%
(b)~\raisebox{-0.9\height}{\includegraphics[width=0.78\textwidth]{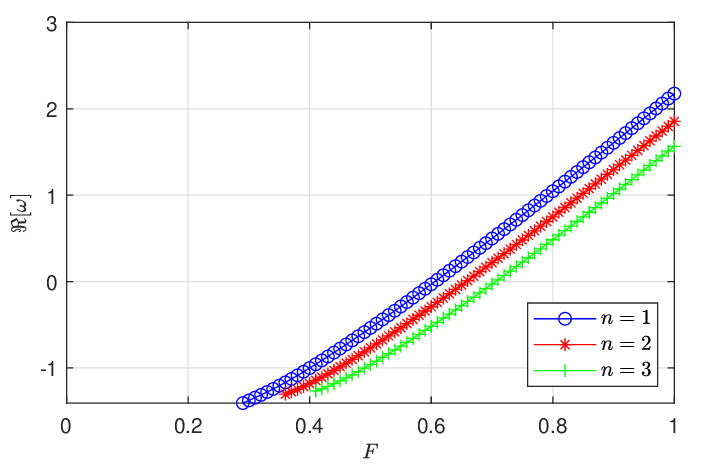}}\par%
\caption{Imaginary (top) and real (bottom) parts of the eigenvalue $\omega_{mn}$ as function of the Froude number $F$ for $m = 7$ and varying $n$.}%
\label{fig:realimag_Vs_F_m7_n123}%
\end{figure}

To look more closely at the trend in the eigensolutions, let us first denote by $n$ an integer representing the number of peaks in radial direction of the modulus of the pressure eigenfunctions, as displayed in figure \ref{fig:peaks_eigenmodes}.
In this way each eigensolution will be indexed by both $m$ and $n$, with corresponding eigenvalues $\omega = \omega_{mn}$.
In figure~\ref{fig:realimag_Vs_F_m7_n123}\textit{a},
the two types of modes can be seen: radiating modes with a significant negative $\Imag(\omega)$, and trapped modes having an almost null $\Imag(\omega)$.
There is no sharp transition between radiating and trapped modes. Accordingly, here we set an arbitrary threshold to separate the two sets of modes by considering a mode to be trapped when its eigenvalue has an imaginary part smaller than $10^{-5}$ in modulus (although our results are relatively insensitive to this threshold; see figure~\ref{fig:rad_trapped_contour} discussed below).
Hence, while the trapped modes considered here are almost neutrally stable, they have eigenvalues with small negative imaginary part and so radiate very slightly. 
Interestingly, none of the modes computed here are observed to become linearly unstable, and we hypothesise that the base solution is at most marginally stable, but not unstable, to linear perturbations of this type at large Froude number.
This stability can be viewed as a manifestation of the robustness of the Lamb--Oseen ~\citep{fabre-2006}, even in the presence of a free surface. 

Another interesting phenomenon concerns the propagation direction of the surface waves with respect to the rotation of the base vortex flow as its rotation rate is varied. This is shown in figure~\ref{fig:realimag_Vs_F_m7_n123}\textit{b}, which shows how the real part of the eigenvalues varies with the Froude number, again for $m = 7$. 
For $\Real(\omega) < 0$, waves rotate opposite to the base flow (counter-rotating waves), while for $\Real(\omega) > 0$, waves rotate in the same direction as the base flow (co-rotating waves). It is clear that for each $n$, there is a value of Froude number separating counter-rotating from co-rotating surface waves.

\subsection{Far-field behaviour and causality}\label{sec:causality}

\begin{figure}
    \centering
    (a)\raisebox{-0.9\height}{\includegraphics[width=0.46\textwidth]{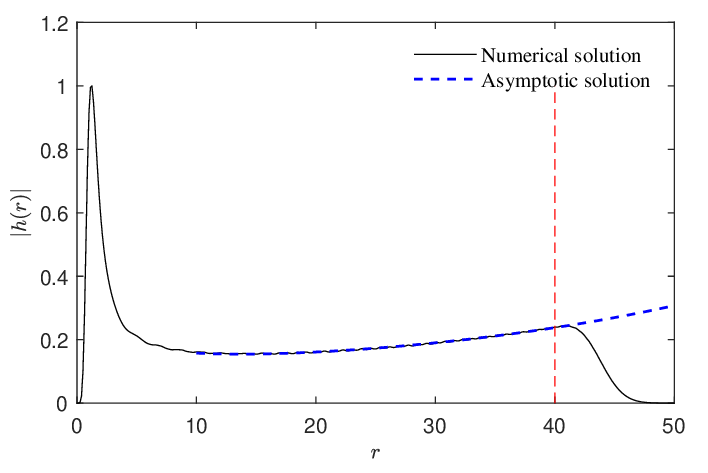}}
    (b)\raisebox{-0.9\height}{\includegraphics[width=0.46\textwidth]{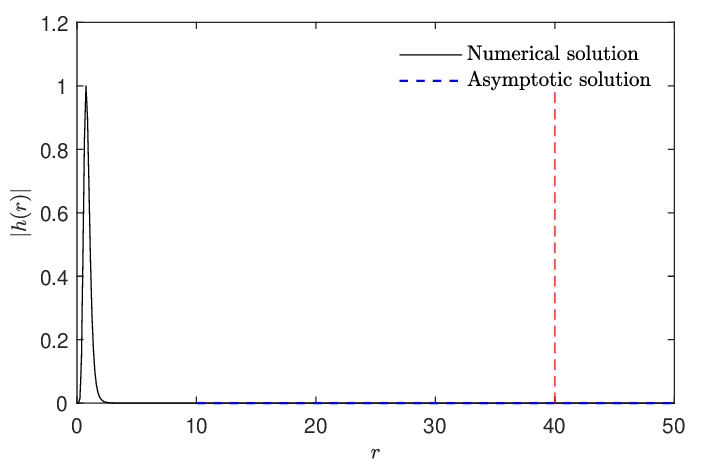}}
    \caption{Comparison of asymptotic and numerical solutions with parameters $m = 7$ and $h_{\infty} = 5$, with $R_c = 40$ and $\bar{\xi} = 10$. (a) A damped radiating mode with $F = 0.3$, giving $\omega=-1.38 - 0.013\I$.  (b) A trapped mode with $F = 0.5$, giving $\omega = -0.54 - 6\times 10^{-10}\I$.  Solid black lines denote the numerical eigenmode, and blue dashed lines denote the far-field asymptotic solution obtained from the deep-water dispersion relation~\eqref{equ:asymp-farfield}. The vertical red dashed lines indicate the edge of the numerical damping region.}
    \label{fig:num_Vs_asympt_solutoins_radiation_BC}
\end{figure}

In this section, we investigate the behaviour of our numerical solutions in the far field, and show that they have the correct behaviour expected of causal outgoing waves.
Away from the vortex core, in the far field, the behaviour of surface waves is well understood.
We first consider the behaviour of deep-water waves in the far-field~\citep[see, e.g.][p. 371]{batchelor}
\begin{align}
h &\propto r^{-1/2}\exp\{\pm\I kr\} &
&\text{with the dispersion relation}\quad \omega^2=k.
\label{equ:asymp-farfield}
\end{align}
For a complex frequency $\omega = \omega_r + \I\omega_i$, the dispersion relation gives $\Imag(k)=2\omega_r\omega_i$.  Causality~(\citealp{briggs-1964,bers-1983}; \citealp[][pp.~17--18]{berti-2009}; \citealp{ledizes+billant-2009,riedinger+etal-2010}) implies that the outgoing wave must decay when $\omega_i>0$, so we must take $\pm = \mathrm{sgn}(\omega_r)$.  Thus, for temporally damped modes when $\omega_i<0$ we expect spatially growing behaviour in the far field.  This growth in the far field is found in our numerical solutions, as shown in figure~\ref{fig:num_Vs_asympt_solutoins_radiation_BC}(a). This behaviour is also seen as $\omega_i \to 0$, but only becomes apparent at extremely large radii (e.g.~$r= 10^{10}$ gives $|h(r)| = 0.01$ in figure~\ref{fig:num_Vs_asympt_solutoins_radiation_BC}(b)), hence our assertion that
eigenmodes with sufficiently small $\omega_i$ can be deemed trapped. 

The growth of eigenmodes in the far field explains why our numerical method breaks down for radiating modes with large temporal decay rate observed at small $F$.
When the temporal decay rate $-\omega_i$ is sufficiently large, 
the spatial growth of the eigenmode overcomes the artificial numerical damping in the damping region, and our numerical non-reflecting boundary condition breaks down.  This results in $\omega$ becoming dependent on the amount of damping in the damping region, and the numerical eigenvalue is automatically removed as being unresolved as previously described in section~\ref{sec:spurious-reflection}.
A similar result has been obtained in \citet{oliveira+cardoso+crispino} studying a model wave equation describing the vortex-waves interaction in the shallow-water limit.
We leave to future studies the possibility of tracking the trend of the spectrum curve as the Froude number goes to zero. It should be noted, that modes having a large negative decay rate in time do not play a relevant role in the dynamics of the system.

\subsection{Extension to other azimuthal wavenumbers.}

The results described above for $m = 7$ are found to be typical at larger azimuthal wavenumbers.
\begin{figure}%
\centering%
{\includegraphics[]{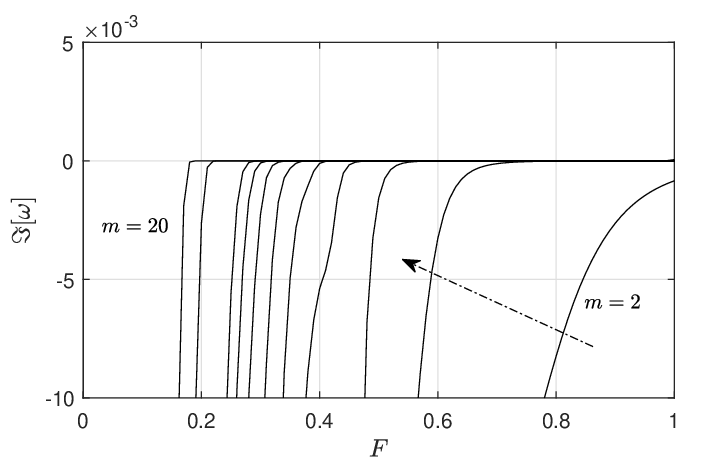}}
\caption{Imaginary part of the eigenvalues as functions of $F$ for $m = 2, 3, 4, 5, 6, 7, 8, 9, 10, 15, 20$, with the arrow indicating the direction of increasing $m$. }%
\label{fig:imag_Vs_F_m789101520_n1}%
\end{figure}%
Figure~\ref{fig:imag_Vs_F_m789101520_n1} shows the trend of the imaginary part of the eigenvalues as a function of the Froude number.
The results are qualitatively similar to the $m=7$ case. Eigenvalue branches shift to lower $F$ with increasing $m$, thus shifting to lower $F$ the value of $F$ separating radiating and trapped modes. 
This suggests that for very high azimuthal wavenumber perturbations we expect to get radiating eigenmodes at lower and lower Froude numbers, as shown in figure~\ref{fig:free_surface_above_m15_F0.19_m20_F0.16}
\begin{figure}%
\centering%
  \includegraphics[width=0.49\textwidth]{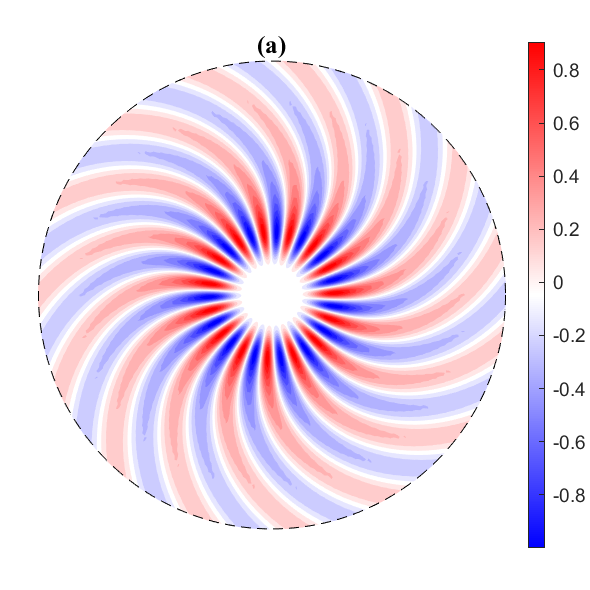}%
\includegraphics[width=0.49\textwidth]{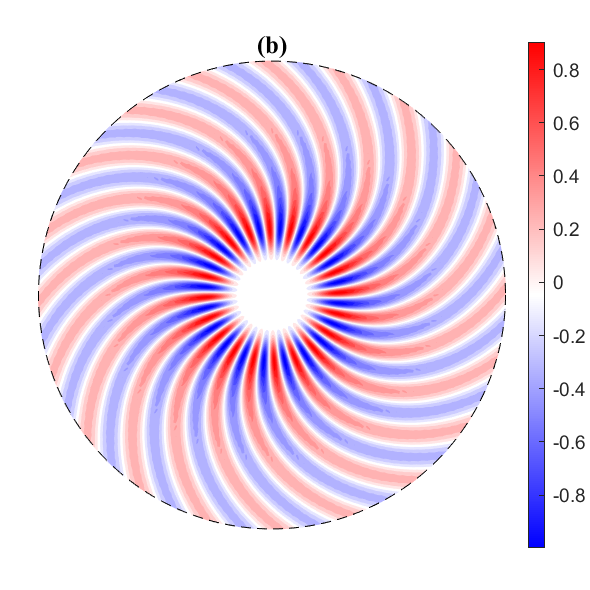}%
\par%
\caption{Shape of the free surface height $h(r, \theta, t=0) = \Real[\phi(r, h_{0}(r))\exp\{\I m\theta\}]$ in two high azimuthal wavenumber cases. (a): $m = 15, F = 0.19$. (b): $m = 20, F = 0.16$.}%
\label{fig:free_surface_above_m15_F0.19_m20_F0.16}%
\end{figure}%
for the example cases of $m = 15$ and $m = 20$.

\begin{figure}%
\centering%
    \includegraphics[]{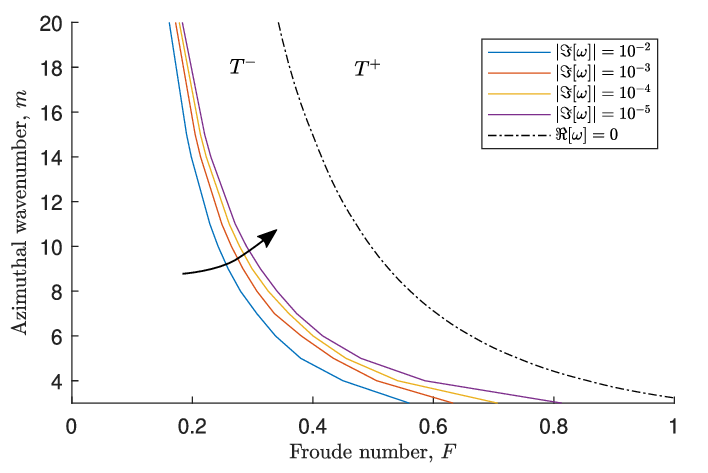}%
\caption{Regimes of radiating (large $|\Imag(\omega)|$) and trapped (small $|\Imag(\omega)|$) modes for different azimuthal wavenumbers $m$ and Froude numbers $F$.  Solid lines are contours of $|\Imag(\omega)|$, with the arrow indicating the direction of more perfectly trapped behaviour. The dash-dot line separates counter-rotating modes ($T^{-}$) with $\Real(\omega)<0$ from co-rotating modes ($T^{+}$) with $\Real(\omega)>0$.}%
\label{fig:rad_trapped_contour}%
\end{figure}

We have computed the range of Froude numbers in which modes radiate and transition towards a neutrally stable state for different values of $m$. This is summarized in figure~\ref{fig:rad_trapped_contour}
showing the classes of solutions previously described over a wide range of azimuthal orders $m$ and Froude numbers $F$.  The contour lines plotted separate the regions of parameter space where waves are radiating and where waves are trapped, with the arrow indicating the direction of better trapping.  While the individual contours range from $\Imag(\omega)=10^{-2}$ to $\Imag(\omega) = 10^{-5}$, their close spacing shows that the exact threshold value of $\Imag(\omega)$ separating trapped and radiating behaviour is not that important, with all contours giving a similar boundary.  Also plotted in figure~\ref{fig:rad_trapped_contour} is a dashed line showing $\Real(\omega)=0$, which is the boundary between counter-rotating and co-rotating trapped waves. Two notable features of the plot as $m$ is decreased are the shift of contours to larger $F$ and the widening of the separation between contours. Both of these features are consistent with figure~\ref{fig:imag_Vs_F_m789101520_n1}, where one sees not only a shift in the eigenvalue curves with $m$, but also a steepening of the transition between radiating and trapped modes with increasing $m$.

While this section considers a wide range of values of $m \geq 3$, we find $m=0, 1$ and $2$ to be dominated by inertial waves rather than surface waves, which are discussed further in section~\ref{sec:inertial}.

\subsection{Effect of the free-surface height on the eigenmodes}

In this subsection we investigated the effects associated to a variation in the fluid depth, exploring the regime between shallow and deep water. We again focus on modes with azimuthal modenumber $m = 7$ and show how the structure of eigenmodes and corresponding eigenvalues vary with fluid depth, and hence how the transition between radiating and trapped modes changes with depth.

\begin{figure}%
\centering%
    \includegraphics[]{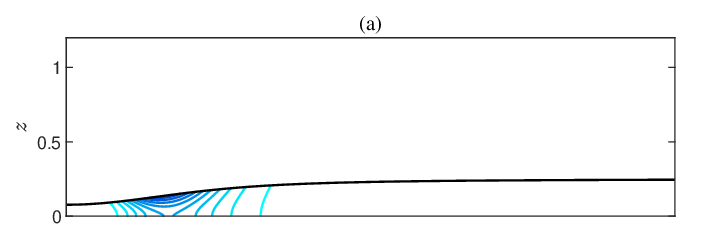}\par%
    \includegraphics[]{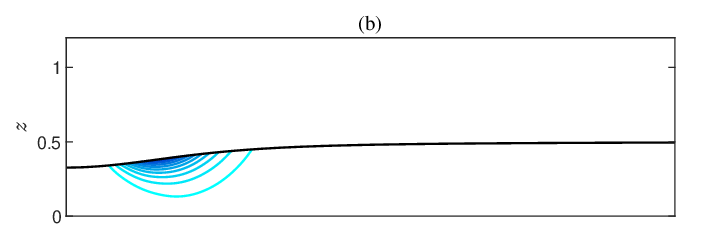}\par%
    \includegraphics[]{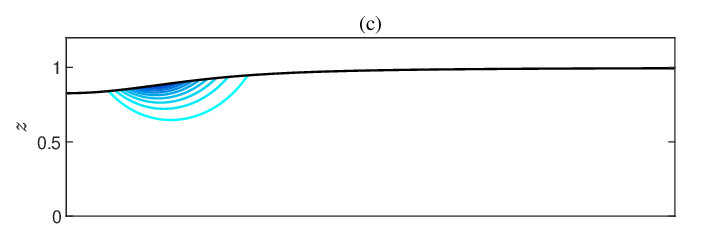}\par%
    \includegraphics[]{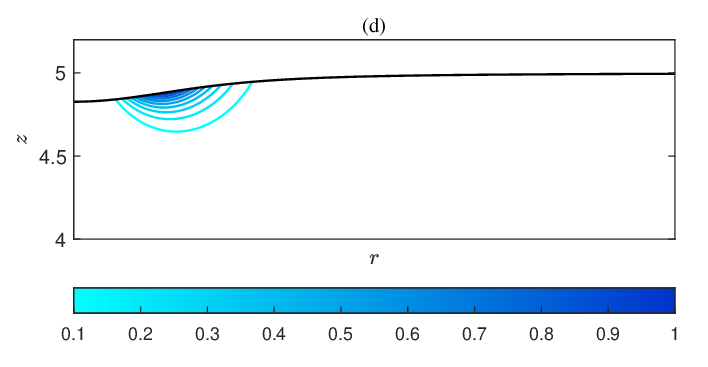}%
    \caption{Effect of the fluid height variation on a single peak ($n = 1$) trapped pressure mode computed for $m = 7$, and $F = 0.5$. (a): $\omega = -0.3058$, $h_{\infty} = 0.25$. 
    (b): $\omega = -0.5316$, $h_{\infty} = 0.5$
    (c): $\omega = -0.5366$, $h_{\infty} = 1$
    (d): $\omega = -0.5370$, $h_{\infty} = 5$.
    The imaginary part of the eigenvalues reported here is at least of order $10^{-9}$, hence can be considered as purely real.}%
\label{fig:effect_hinf_mode_F05_n1}%
\end{figure}

\begin{figure}%
\centering%
    \includegraphics[]{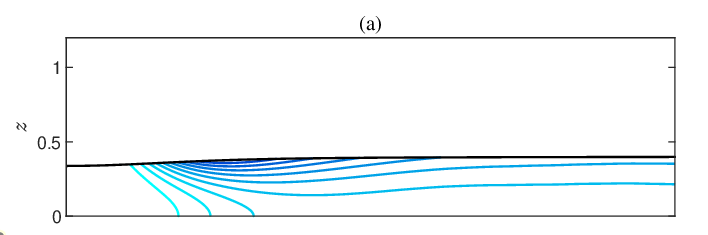}\par%
    \includegraphics[]{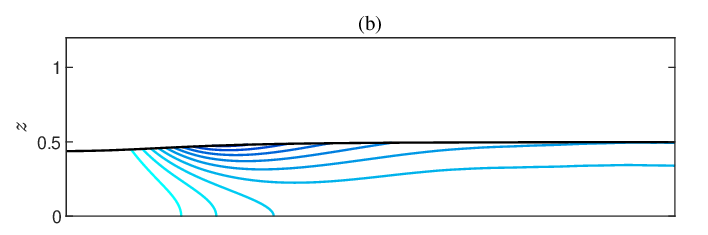}\par%
    \includegraphics[]{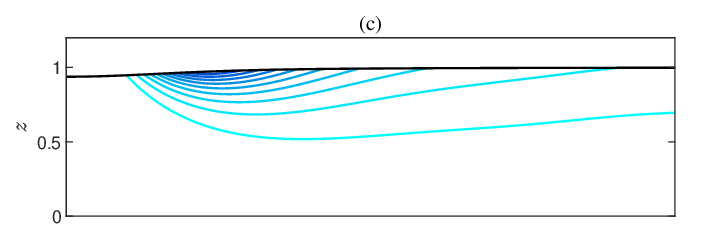}\par%
    \includegraphics[]{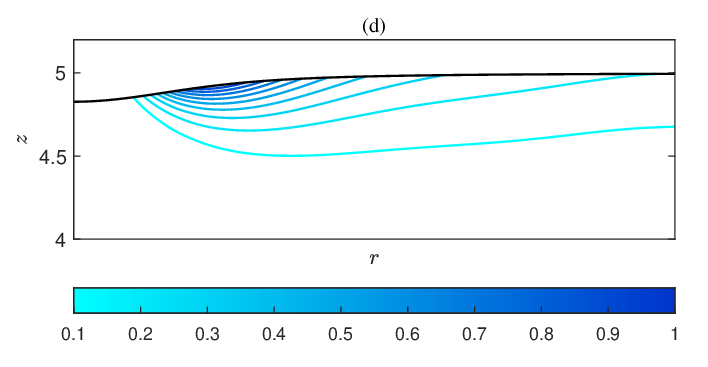}%
    \caption{Effect of the fluid height variation on a radiating pressure mode computed for $m = 7$, and $F = 0.3$. (a): $\omega = -1.3371 - 0.0530\I$, $h_{\infty} = 0.4$. 
    (b): $\omega = -1.3633 - 0.0320\I$, $h_{\infty} = 0.5$
    (c): $\omega = -1.3772 - 0.0140\I$, $h_{\infty} = 1$
    (d): $\omega = -1.3770 - 0.0138\I$, $h_{\infty} = 5$.}%
\label{fig:effect_hinf_mode_F03}%
\end{figure}

\begin{figure}%
\centering%
    \includegraphics[]{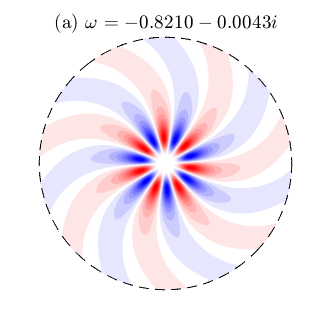}%
    \includegraphics[]{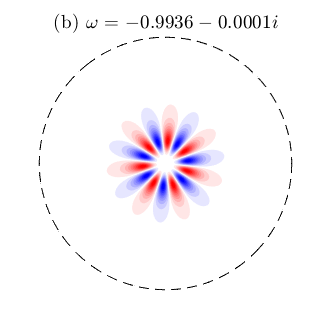}%
    \par%
    \includegraphics[]{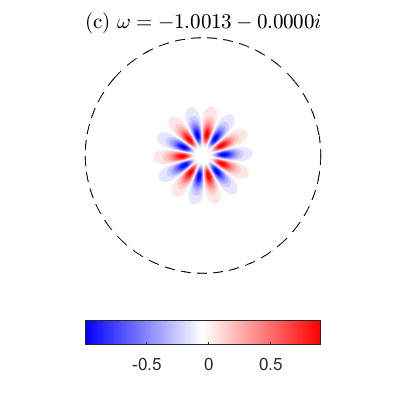}%
    \caption{Free surface height for $m = 7$, $F = 0.4$ and different fluid depths. (a): $h_{\infty} = 0.25$. (b): $h_{\infty} = 0.5$. (c): $h_{\infty} = 1$.}%
\label{fig:effect_hinf_modes_F04}%
\end{figure}

\begin{figure}%
\centering%
    \includegraphics[]{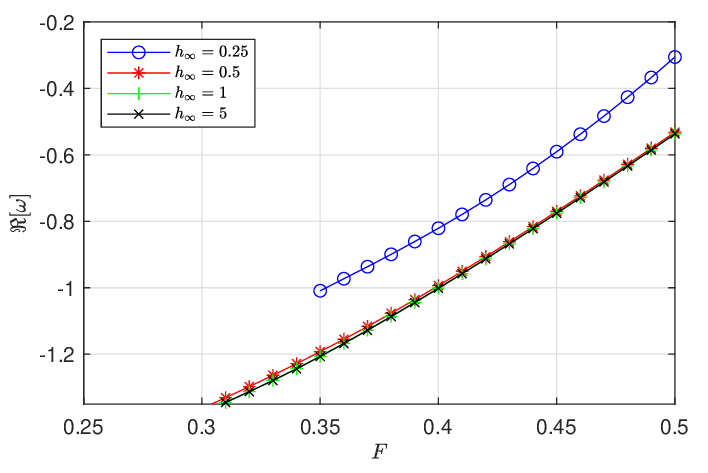}%
    \par
    \includegraphics[]{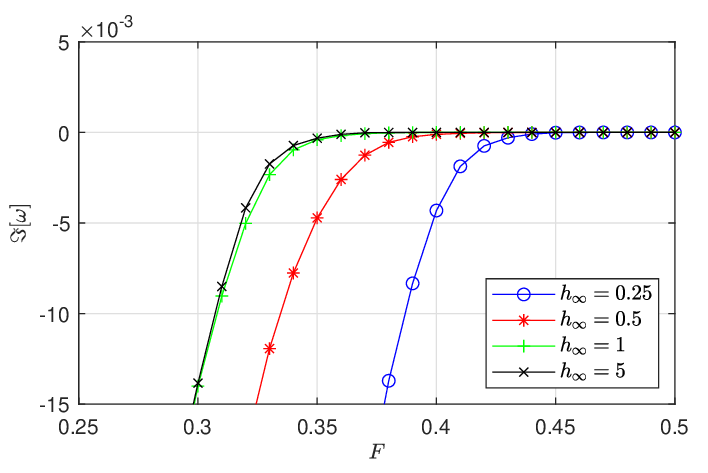}%
    \caption{Effect of the fluid height variation on the spectrum curves for the least damped set of eigenmodes obtained for $m = 7$.}%
\label{fig:spectrum_curves_different_heights}%
\end{figure}

The variation in the structure of the surface gravity eigenmodes with fluid depth $h_{\infty}$ is displayed in figure \ref{fig:effect_hinf_mode_F05_n1}
for a typical trapped mode computed at Froude number $F = 0.5$. The same type of analysis is carried out for a radiating mode at $F = 0.3$, whose spatial structures are shown in figure \ref{fig:effect_hinf_mode_F03}.
Finally, in figure \ref{fig:effect_hinf_modes_F04},
the transition between radiating and trapped mode is shown as the fluid depth is varied between $h_{\infty} = 0.25, 0.5, 1$. 
Figure~\ref{fig:spectrum_curves_different_heights}
shows the variation in eigenvalue as the Froude number is varied, for different fluid depths.  The trend in $\Imag(\omega)$ shown suggests that more shallow systems require a faster rotating vortex in order to get the same wave trapping as a deeper water configuration.  The variation with depth of $\Real(\omega)$ is less significant.  For $h_{\infty} = 1$, the eigenvalues are essentially identical to those computed for even higher depths, for example, in this case $h_{\infty} = 5$.
Hence, in this regard, for a trapped mode a fluid with $h_{\infty} \in [0.5, 1]$ can already be considered deep water.

\subsection{Inertial modes}
\label{sec:inertial}

While the focus of this study is surface gravity waves, our numerics finds modal solutions to the governing equations~\eqref{equ:linearized} without any assumption about the modes being surface gravity waves.  By way of contrast, therefore, in this section we briefly discuss another type of mode, namely inertial modes. These modes persist even at low rotation rates, and are characterised by being neutrally stable as well as being concentrated within the vortex core and not being localized close to the free surface. Our results are therefore similar to the numerical results presented by \citet{mougel-2015} for a solid-body rotation, since the Lamb--Oseen vortex is very close to solid-body rotation close to the centre.
\begin{figure}%
\centering%
    \includegraphics[width=6cm]{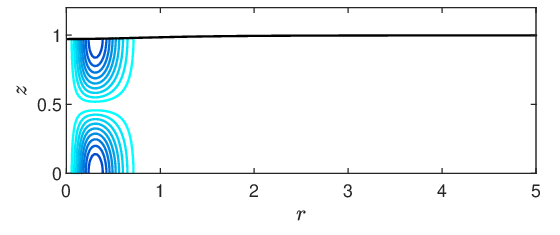}%
    \includegraphics[width=6cm]{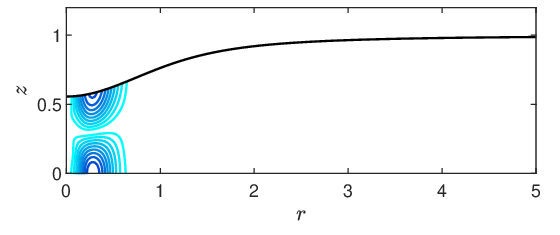}%
    \par%
    \includegraphics[]{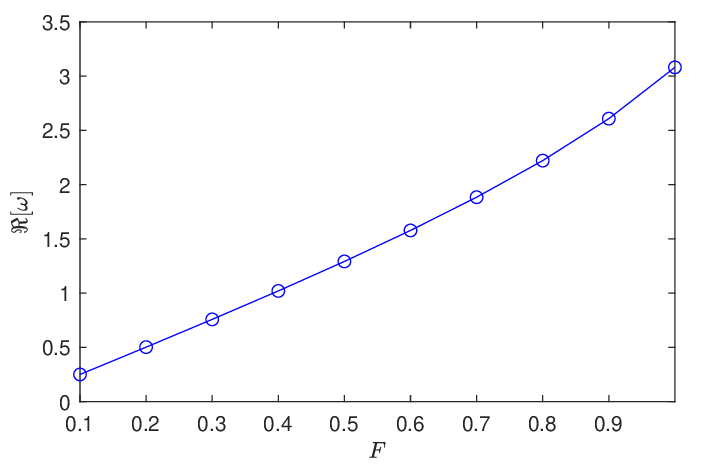}%
    \caption{Inertial modes for $m = 2$, $h_{\infty} = 1$ and $F = 0.2$ (left); $F = 0.8$ (right). Eigenvalues curve as function of the Froude number for the same parameters (bottom figure). }%
\label{fig:inertial_modes}%
\end{figure}%
Figure~\ref{fig:inertial_modes} shows two inertial modes computed for $m = 2$, $h_{\infty} = 1$, and $F = 0.2, 0.8$. The corresponding eigenvalues are purely real, and are also plotted as function of the Froude number in figure~\ref{fig:inertial_modes}.
%========================================================
%
% END RESULTS
%
% ========================================================
\section{Conclusions and future developments}\label{sec:conclusions}

% Summary
We have considered the linear response to small perturbations of a free surface Lamb--Oseen vortex flow in a laterally unbounded domain.
Throughout this work, we have assumed a harmonic dependence $\exp\{-\I\omega t + \I m\theta\}$ (equation~\ref{equ:modal}) and looked for modal solutions of this form.
We restricted our focus to surface waves, as it was these waves that were seen in the motivating experiments; we note that the numerical procedure described here also calculates other types of modes which we have not investigated in depth here, and which may in some circumstances be the least damped modes, especially at low Froude numbers.
%

% Trapped vs Radiating modes
Our study classified surface waves into two distinct classes: radiating modes and trapped modes.  While these have previously been seen for shallow water vortices with inflow (such as a plughole vortex), here we have shown that they arise in a fully 3D non-shallow-water problem without the need for an inflow to help trap the modes. Radiating modes are dissipative, and have a spatial structure which extends in the horizontal plane, therefore behaving as radially-travelling waves.
By contrast, trapped modes, by virtue of their near zero growth rate, persist for long times with little dissipation and remain localized within the core region of the vortex and hence could be expected to be observed on the surface of a rotating vortex after all damped modes have dissipated.
The spatio-temporal behaviour of these modes resembles that of a radially-standing wave instead of a radially-travelling wave. Our numerical predictions show the appearance of trapped modes provided the Froude number $F$ (the dimensionless rotation rate) is above a threshold value, with the threshold depending on the azimuthal mode number $m$ (as summarized in figure~\ref{fig:rad_trapped_contour}). Our computations also suggest that trapped modes will asymptotically approach a neutrally stable state in the limit of large Froude number without becoming unstable, although the exact asymptotic scaling of $\Imag(\omega)$ for large $F$ was not obvious from the numerical analysis carried out here.  
Interestingly, our numerics finds no unstable wave modes with $\Imag(\omega)>0$, unlike in either bounded vortex flows~\citep{tophoj+others,mougel-2014,mougel-2017} or unbounded shallow water vortex flows~\citep{ford}, both of which possess unstable modes with $\Imag(\omega)>0$. (As shown in appendix~\ref{appendix:code_validation_SB_rotation}, our numerics reproduces such instabilities, and thus would be expected to find instabilities in the present case if they existed.)
Finally, we have confirmed what has been seen in the initial motivating experiment of the pool, namely that stable trapped surface wave modes exist and can propagate with or against the base swirling flow (as also summarized in figure~\ref{fig:rad_trapped_contour}).

% The critical layer
Our assumed harmonic dependence $\exp\{-\I\omega t + \I m\theta\}$ (equation~\ref{equ:modal}) means modes with $\Imag(\omega)>0$ are unambiguously unstable and grow exponentially in time, while modes with $\Imag(\omega)<0$ are unambiguously stable and decay exponentially in time.  In the edge case with $\Imag(\omega)=0$, a critical layer may be present, defined as a radial location $r=r_c$ for which $D_t = -\I\omega + \I mF\Omega_0(r_c) = 0$.  This possibility has been neglected here.  The critical layer is related to the continuous spectrum of the differential operator, and analysing the critical layer requires a different analysis and numerical technique: the temporal Fourier transform implicit in the harmonic dependence must be inverted, and the singular point $r=r_c$ must be avoided by complex deformation of the solution into the complex $r$-plane, with the correct choice of deformation to maintain causality~\citep[such as has been done for aeroacoustic waves; e.g.][]{king+brambley-2022}.  Despite the critical layer being present only for $\Imag(\omega)=0$, it may result in a solution with algebraic growth in time.  Consequently, the lack of solutions with $\Imag(\omega)>0$ does not prove the stability of perturbations to a Lamb--Oseen vortex with a free surface, but only the lack of unstable solutions with exponential growth in time.  However, we note that \citet{fabre-2006} found that the Lamb--Oseen vortex without a free surface is stable to axial-invariant perturbations as disturbances related to the critical layer are necessarily damped in that case.

% Comment on the physical mechanism for the modes
Based on the results presented here, we may speculate on the underlying physical mechanism causing this trapping behaviour.  One likely candidate might have been the non-uniform shape of the free surface, with trapped modes appearing to be localized at a radius close to the largest gradient of the undeformed free surface.  However, in appendix~\ref{appendix:results_without_free_surface_deformation} results are shown for numerical simulations with an (artificial) flat free surface, and almost identical results are obtained, suggesting that, at least for the parameters investigated here, the free surface shape is less important than the base flow.  We hypothesise, therefore, that it is the refraction of the surface gravity waves by the base flow, causing the waves to curve around the vortex with exactly the curvature needed to remain trapped, that is the underlying physical mechanism.  This would explain why the non-shallow-water assumption is important here, as the dispersive nature of the waves in this case means a frequency can be chosen to obtain the particular wave group velocities needed for wave trapping, where as in the shallow water case the waves are nondispersive and have a prescribed group velocity independent of their frequency.

% Note about nonreflecting boundary conditions
To numerically simulate a (horizontally-)unbounded fluid on a bounded numerical domain, a far-field nonreflecting boundary condition or buffer region is needed.  Here, a novel additional term is introduced into the governing equations, to provide damping of the surface waves in the buffer region only. This method has proved more accurate than any non-reflecting boundary condition we implemented while remaining computationally viable. Indeed, introduction of additional unknowns in the mathematical formulation was not needed, thus overcoming the main drawback of PML methods. Furthermore, given that the background vortex flow vanishes in the far-field, the same absorbing layer formulation can be employed with other similar vortex distributions.

% 2D numerics
The numerical expense of our eigenvalue problem (which is a two-dimensional spatial problem involving both $r$ and $z$ coordinates) might be reduced by investigating a one-dimensional approximation along the radial coordinate only; for example in either the deep- or shallow-water limits. Moreover, new methods and formulations for the imposition of a non-reflecting boundary condition in the far-field would certainly lead to a saving in the computational time as it will avoid, for example, the need to numerically resolve the unphysical buffer region.

% Future work
Due to the computational expense, we have left to future studies the extension of our parametric study of waves to extreme Froude numbers. In particular it would be interesting to investigate the trend of modes for very small and very large Froude numbers. Also, consideration of lower or higher aspect ratios $h_{\infty}$ could be investigated.
The assumption of modal solutions with a harmonic dependence $\exp\{-\I\omega t + \I m\theta\}$ (equation~\ref{equ:modal}) could also be relaxed, for example through time-domain simulation, which would allow investigation into the critical layer~\citep[e.g. as in][]{riedinger+etal-2010,king+brambley-2022}.
Even with the assumption of harmonic dependence $\exp\{-\I\omega t + \I m\theta\}$, a different but related problem would be to investigate the scattering of an incoming wave encountering the vortex, which would require a different far-field boundary condition to introduce the incoming wave as well as to allow outgoing waves to propagate through the far-field boundary without reflection.  Finally,
our model also neglects both nonlinear and surface tension effects; while this is justified for the swimming pool application we model here, these assumptions break down for either large amplitudes or short wavelengths.
Therefore, it would be interesting to investigate whether their inclusion could lead to an instability of the base vortex flow.

\begin{acknowledgements}
\noindent\textbf{Acknowledgements.} E.~Zuccoli was funded through the Warwick Mathematics Institute Centre for Doctoral Training, and gratefully acknowledges the support of the University of Warwick and the UK Engineering and Physical Sciences Research Council (EPSRC grant EP/W523793/1).
E.J.~Brambley gratefully acknowledges the support of the UK Engineering and Physical Sciences Research Council (EPSRC grant EP/V002929/1).
D.~Barkley gratefully acknowledges supported from the Simons Foundation (Grant No. 662985).
The authors are grateful to J.M.~Skipp for the initial motivation and videos motivating this project.
For the purpose of open access, the authors have applied a Creative Commons Attribution (CC BY) licence to any Author Accepted Manuscript version arising from this submission.

\vspace{1ex}\noindent\textbf{Declaration of Interests}. The authors report no conflict of interest.
\end{acknowledgements}

\appendix

\section{Detailed derivation and analysis of the Numerical Solution}
\label{appendix:numerics}

\subsection{Numerical solution of the eigenvalue problem}

In order to solve~(\ref{equ:linearized},\ref{equ:linearized-bcs},\ref{equ:modified-continuity}), we used a Galerkin spectral method and expand with a combination of Legendre polynomials as basis functions in both the radial and axial coordinate in order to satisfy the Dirichlet boundary conditions. First of all, to account for the shape of the computational domain with a variable surface height $h_0(r)$, we remap the domain from the physical domain $D = [0, R]\times[0, h_{0}(r)]$ to the computational square domain $S = [-1, 1]\times[-1, 1]$ using the non-orthogonal transformation
\begin{align}
x &= \frac{2}{R}r - 1, &
y &= 2\frac{z}{h_{0}(r)} - 1.
\label{change_of_variables}
\end{align} 
In this new domain $S$, the problem becomes
\begin{subequations}\begin{gather}
\I m F \Omega_{0}(x)u - 2 F \Omega_{0}(x) v + \frac{2}{R}\!\left[\frac{\partial\phi}{\partial x} - \frac{h'_{0}(x)}{h_0(x)}(y + 1)\frac{\partial\phi}{\partial y}\right]\! = \I\omega u, \label{equ:x_momentum_xy_vars}\\
\I m F \Omega_{0}(x)v + \frac{2F}{R(x+1)}\frac{d}{dx}\left[ (x+1)U_{0}(x) \right]  u + \frac{2\I m}{R(x + 1)}\phi = \I\omega v, \\
\I m F \Omega_{0}(x)w + \frac{2}{h_0(x)}\frac{\partial \phi}{\partial y} = \I\omega w, \\[1ex]
\begin{aligned}
\frac{2}{R(x+1)}\frac{\partial}{\partial x}\big[(x+1)u\big] &+ \frac{2}{R}\frac{h'_{0}(x)}{h_{0}(x)}u + \frac{2\I m}{R(x + 1)}v \\&+ \frac{2}{h_0(x)}\frac{\partial}{\partial y}\!\left[ w - \frac{h'_{0}(x)}{R}(y+1)u \right]\! + \xi(x)\phi = 0,
\end{aligned}\\[1ex]
\I m F \Omega_{0}(x)\phi(x, 1) + \frac{2h_0'(x)}{R}u(x, 1) - w(x,1) = \I\omega\phi(x,1), \\[1ex]
w(x, -1) = 0,
\qquad\quad
\phi(1,y) = 0,
\qquad\quad
|\vect{u}(-1, y)| < \infty,
\qquad\quad
|\phi(-1, y)|  < \infty.
\end{gather}\label{equ:equations_remapped}\end{subequations}
For $R_c \leq r \leq R$, we include a damping layer with a nonzero $\xi(r)$.  Here, we choose a simple form for $\xi(r)$, which we demonstrate in section~\ref{sec:convergence} works well for the parameters we consider here:
\begin{equation}
\xi(r) = 
\begin{cases}
0, & r < R_{c}, \\
\bar{\xi}\left(\dfrac{r - R_{c}}{R - R_{c}}\right)^{2}, & R_{c} \le r \le R.
\end{cases}
\label{equ:damping_function}
\end{equation}
Since $\xi(r)$ is continuously differentiable at $r=R_{c}$, no special transmission conditions are needed there~\citep[][p.~46]{sim-2010}.

We next obtain the weak discretized formulation of this problem.
The basis functions are chosen such that they identically satisfy the homogeneous Dirichlet boundary data. Also, the specific expansion of each variable in terms of basis functions will depend on the value of the azimuthal wavenumber, as this is associated with a singularity at the origin for the pressure and the aximuthal velocity component. In the following we make the distinction between the case of axisymmetric perturbations ($m = 0$), and generic non-axisymmetric perturbations ($m \neq 0$).

\subsubsection{Discrete problem for non-axisymmetric perturbations}

In the general case of a nonzero value of $m$, the weak formulation of the eigenvalue problem can be obtained in the following way. First of all, we notice the presence of singular terms at $r=0$ for $u, v, \phi$ so we will require these function to be null at the origin. Consequently, even $w$ will be so. Thus, we look at the unknowns in the following spaces:
\begin{subequations}\begin{gather}
\label{spaces_generic_m}
[u, v, \phi] \in V_{H}(S) = \{ (u, v, \phi) \in H^{1}(S): (u, v, \phi) = 0, \quad at \quad r = 0 \}, \\
w \in V_{v}(S) = \{ w \in H^{1}(S): w = 0, \quad at \quad r = 0, \quad z = 0 \},
\end{gather}\end{subequations}
where $H^{1}(S)$ is the usual Sobolev space
\citep[ch. 2]{quarteroni-2009}.
We multiply the azimuthal component of the momentum equation and the continuity equation by $(x+1)$. Then, by multiplying each of the equations by suitable test functions $(v_{x}, v_{t}, v_{y}, q)$ in the same space as the corresponding unknowns, after integrating over the square $S$ and exploiting the boundary conditions, the weak formulation of the differential problem reads: find $([u, v, \phi], w) \in V_{H}(S) \times V_{v}(S)$ such that $\forall ([v_{x}, v_{t}, q], v_{y}) \in V_{H}(S) \times V_{v}(S)$ the following holds:
\begin{subequations}\begin{gather}
\int_{S} v_{x}\I m F \Omega_{0}(x)u - 
\int_{S} v_{x} 2 F \Omega_{0}(x) v - 
\int_{S} \frac{2}{R}\frac{\partial v_{x}}{\partial x}\phi - 
\int_{S} v_{x} \frac{2}{R}\frac{h'_{0}(x)}{h_0(x)}(y + 1)\frac{\partial\phi}{\partial y} = \I\omega \int_{S} v_{x} u, \\
\int_{S} v_{t}\left[\I m F \Omega_{0}(x)(x+1)v + \frac{2F}{R}\!\left[(x+1)U_{0}(x)\right]\!u + \frac{2\I m}{R}\phi\right] = \I\omega\int_{S} v_{t} (x+1) v, \\
\int_{S} v_{y}\left[\I m F \Omega_{0}(x)w + \frac{2}{h_0(x)}\frac{\partial \phi}{\partial y}\right] = \I\omega \int_{S} v_{y} w, \\
\begin{aligned}
\int_{S} &\frac{2}{R} q \frac{\partial}{\partial x}\left[(x+1)u\right] +
\int_{S} \frac{2}{R}\frac{h'_{0}(x)}{h_{0}(x)}(x+1) \left[q u + \frac{\partial q}{\partial y} u\right]
+ \int_{S} \frac{2\I m}{R}q v
\\&
-\int_{S} \frac{2}{h_{0}(x)}(x+1)\frac{\partial q}{\partial y} w 
+ \int_{x=-1}^{x=1} \frac{2\I m F (x+1)\Omega_{0}(x)}{h_{0}(x)}q(x, 1)\phi(x, 1)
\\&= 
\I \omega \int_{x=-1}^{x=1} \frac{2(x+1)}{h_{0}(x)} q(x, 1)\phi(x, 1).
\end{aligned}
\end{gather}%
\label{equ:weak_form_explicit}%
\end{subequations}%
Let us define the bilinear forms $\mathcal{A}: V_{H} \times V_{v} \rightarrow \mathbb{R}$ and $\mathcal{B}: V_{H} \times V_{v} \rightarrow \mathbb{R}$ such that the generalised eigenvalue problem above can be compactly written as: find $([u, v, \phi], w) \in V_{H} \times V_{v}$ such that
\begin{equation}
\begin{aligned}
    \mathcal{A}([u, v, w, \phi], [v_{x}, v_{t}, v_{y}, q]) = \omega \mathcal{B}([u, v, w, \phi], [v_{x}, v_{t}, v_{y}, q]), \quad \forall ([v_{x}, v_{t}, q], v_{y}) \in V_{H} \times V_{v}
\end{aligned}
\label{equ:weak_form_elegant}
\end{equation}
At this point, in order to discretize the problem, we need to expand the unknowns in terms of proper basis functions. Such basis functions are taken in such a way the homogeneous Dirichlet boundary conditions are automatically satisfied in both the axial and radial coordinates.
In particular, we define the following set of polynomials, $P_{n}^{\star}(x)$ as follows:
\begin{equation}
P_{n}^{\star}(x) = P_{n}(x) + P_{n-1}(x), \quad n\ge 1,
\label{equ:modified_legendre_polynomials}
\end{equation}
where $P_{n}(x)$ are standard Legendre polynomials. 

Thanks to the above definition, a Dirichlet boundary condition at $x = -1$ is automatically satisfied. Therefore, we expand the velocity components and the pressure as
\begin{equation}
\begin{aligned} &
[u, v, \phi](x, y) = \sum_{i=1}^{N_x}\sum_{j=1}^{N_y}[u_{ij}, v_{ij}, \phi_{ij}]P_{i}^{\star}(x)P_j(y), \\ &
w(x, y) = \sum_{i=1}^{N_x}\sum_{j=1}^{N_y}w_{ij}P_{i}^{\star}(x)P_{j}^{\star}(y).
\end{aligned}
\label{expansion_unknowns_genericm}
\end{equation}
After substituting the expansion into the weak formulation (\ref{equ:weak_form_elegant}), we obtain the final discrete eigenvalue problem, which reads
\begin{equation}
\begin{aligned} &
\mat{A}\vect{w} = \omega \mat{B}\vect{w}, 
\end{aligned}
\label{equ:discrete_evals_problem}
\end{equation}
where $\vect{w} = (u_{ij}, v_{ij}, w_{ij}, \phi_{ij})$ is our array containing the spectral coefficients previously defined, $\mat{A}$ and $\mat{B}$ are the matrices of order $4N_{x}N_{y} \times 4N_{x}N_{y} $coming from the evaluation of the integrals appearing in the weak formulation for each equation.
% ---------------------------------------------------
%
%
%
% ---------------------------------------------------
\subsubsection{Discrete problem for axisymmetric perturbations}

In the axisymmetric case ($m = 0$), the differential problem~\eqref{equ:equations_remapped} over the square $S$
simplifies, and it can be noted that only the radial component of the velocity must go to zero as $r\rightarrow 0$, with the other unknowns allowed to take any finite value. Thus, studying the axisymmetric perturbation problem means to look for the eigensolutions in the following spaces: $u \in V_{H}(S)$, $(v, \phi) \in H^{1}(S)$ and $w \in V_{v0} = \{ w \in H^{1}(S): w = 0, \quad at \quad z = 0 \}$. The weak formulation is then obtained as shown previously. Regarding the disctetization process, we express the four unknowns (and corresponding test functions) as follows
\begin{equation}
\begin{aligned} &
u(x, y) = \sum_{i=1}^{N_x}\sum_{j=1}^{N_y}u_{ij}P_{i}^{\star}(x)P_j(y), \\ &
[v, \phi](x, y) = \sum_{i=1}^{N_x}\sum_{j=1}^{N_y}[v_{ij}, \phi_{ij}]P_{i}(x)P_j(y), \\ &
w(x, y) = \sum_{i=1}^{N_x}\sum_{j=1}^{N_y}w_{ij}P_{i}(x)P_{j}^{\star}(y).
\end{aligned}
\label{expansion_unknowns_axisymmetric}
\end{equation}
After substitution of the expansion above into the weak formulation, the resulting algebraic generalised eigenproblem (\ref{equ:discrete_evals_problem}) is recalled in the case $m = 0$.
% ----------------------------------------------------
%
%
%
% ----------------------------------------------------
\subsection{Spurious numerical modes and resolvedness conditions}
\label{sec:resolvedness}

Solving the discretized eigenvalue problem~\eqref{equ:discrete_evals_problem} is known to give rise to spurious eigenvalues and eigenfunctions. These are numerical artifacts caused by numerical under-resolution of highly-oscillatory modes; that is, eigenfunctions do satisfy the discretized problem but do not approximate solutions of the continuous eigenvalue problem. Following~\citet{brambley+peake-2008} \citep[see also][p.~78--81]{brambley-2007}, we implement two tests to remove such spurious numerical modes.

The first condition requires that the eigenvalues do not move significantly when the numerical discretization is changed.  Suppose eigenvalues $\{\omega_{p}\}$ have been computed using a discretization $(N_x, N_y)$, and suppose $\{\hat{\omega}_{q}\}$ have been computed with a slightly increased resolution, such as $(N_{x}+1, N_{y}+1)$.  The first condition is then given by requiring that, for each eigenvalue $\omega_p$,
\begin{equation}
\min_q|\omega_{p} - \hat{\omega}_{q}| < \text{tol},
\label{resolv_cond_eigenvalues}
\end{equation}
where $\text{tol}$ is a prescribed tolerance. For the results computed in this paper $\text{tol} = 10^{-2}$ was used.

The second resolvedness condition involves the spectral coefficients of the eigenfunctions. In particular, we require the modulus of the spectral coefficients to decay smoothly as the number of polynomials used gets larger. Both the aforementioned conditions have been applied to the pressure eigenfunctions as this is the unknown of most interest because it contains information on the shape of the free surface height.
So, let $\phi_{ij}$ be a generic spectral coefficient, the following condition has been implemented
\begin{equation}
\begin{aligned} &
\frac{\sup_{(i, j) \in \mathcal{B}} |\phi_{ij}|}{\sup_{\mathrm{all}\,i, j} |\phi_{ij}|} < \text{tol}, 
\end{aligned}
\label{resolvedness}
\end{equation}
where $\mathcal{B}$ is a part of the domain in the $N_{x}-N_{y}$ plane, defined as
\begin{equation}
\mathcal{B} = \big\{ (i, j): N_{x}+1 - b_{x} \le i \le N_{x}+1 \text{ or } N_{y}+1 - b_{y} \le j \le N_{y}+1 \big\},
\end{equation}
$\text{tol}$ is a prescribed tolerance on the magnitude of the coefficients and $b_x, b_y$ two prescribed borders widths in the $N_{x}-N_{y}$ plane.
\begin{figure}
\centering%
    \includegraphics[width=0.7\textwidth]{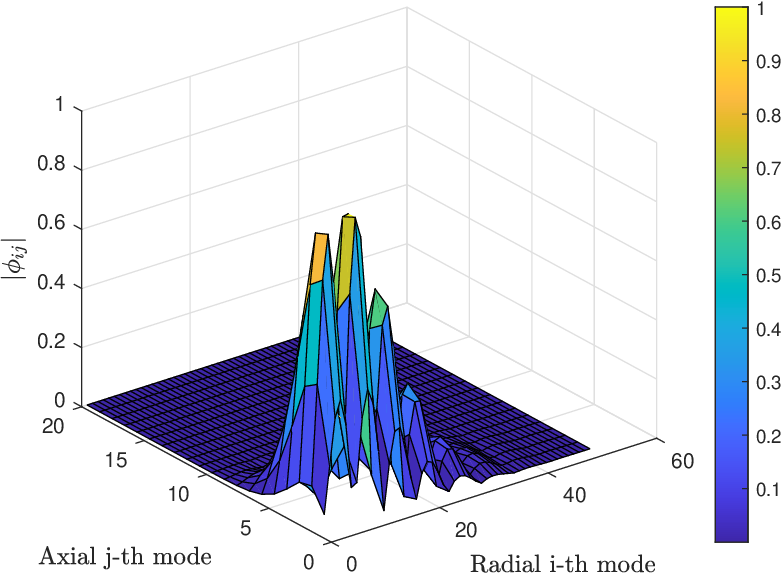}
    \includegraphics[width=0.7\textwidth]{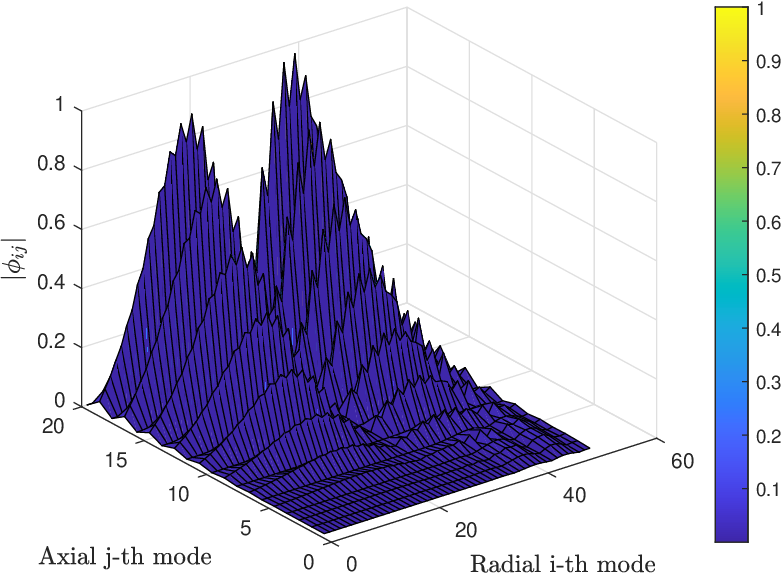}
\par%
\caption{Example of a well-resolved mode (Top) and a poorly-resolved mode (Bottom).}
\label{fig:well_bad_resolved_modes}
\end{figure}
Figure~\ref{fig:well_bad_resolved_modes} compares the magnitude of the coefficients for a well resolved mode and a poorly-resolved mode for $N_{x} = 50$, $N_{y} = 20$, $b_x = 12$, and $b_y = 4$, $\text{tol} = 10^{-1}$.  The well-resolved mode can be seen to have spectral coefficient $|\phi_{ij}|$ decreasing exponentially quickly. By contrast, for the poorly-resolved mode, the spectral coefficients $|\phi_{ij}|$ are oscillatory increasing as function of the number of polynomials used.
% ====================================================
%
% END APPENDIX A ON DETAILED NUMERICAL PROCEDURE.
%
% =====================================================
%
%
% 
% Appendix on damping layer formulation and compressibility term.
%

\section{Damping layer formulation}
\label{appendix:damping_layer_form}

In section~\ref{section:absorbing-layer}, the continuity equation was modified by the introduction of the ``damped compressibility'' term $\xi(r)\phi$ in equation~\eqref{equ:modified-continuity} in order to emulate an infinite domain using our finite computational domain, by damping out disturbances as they reach the unphysical computational boundary at $r=R$.  Here, we justify the inclusion of that term by analogy with the equations of acoustics.

Small (linear) homentropic perturbations $(\vect{u}, p, \rho)$ to a static fluid $(\vect{0}, p_0, \rho_0)$ are governed by the linearized Euler Equations:
\begin{align}
        \rho_{0}\frac{\partial\boldsymbol{u}}{\partial t} + \nabla p &= 0, &
        \frac{1}{c^2}\frac{\partial p}{\partial t} + \rho_0 \nabla\cdot\boldsymbol{u} &= 0, &
        p = c^{2}\rho,
\end{align}
where $c$ is the speed of sound.  These can be combined into a single wave equation for the pressure,
\begin{equation}
        \frac{1}{c^2}\frac{\partial^2 p}{\partial t^2} - \nabla^{2}p = 0.
    \label{equ:wave_eq_pressure}
\end{equation}
Following \citet[][pp. 81-82]{gao+others}, in order to introduce a sponge layer to damp outgoing waves, we add a damping term of the form $\xi(r)\partial_{t}p$ into the wave equation~\eqref{equ:wave_eq_pressure},
\begin{equation}
        \frac{1}{c^2}\frac{\partial^2 p}{\partial t^2} + \xi(r)\frac{\partial p}{\partial t} - \nabla^{2}p = 0.
    \label{equ:wave_eq_pressure_damping}
\end{equation}
This wave equation can be split back into the original physical mass- and momentum-equations in the original physical variables $(\boldsymbol{u}, p, \rho)$ as
\begin{align}
        \rho_{0}\frac{\partial\boldsymbol{u}}{\partial t} + \nabla p &= 0, &
        \frac{1}{c^2}\frac{\partial p}{\partial t} + \xi(r)p + \rho_0 \nabla\cdot\boldsymbol{u} &= 0, &
        p = c^{2}\rho,
    \label{equ:system_far_field_with_damping}
\end{align}
If we now take the incompressible limit $1/c^2 \to 0$ in system~\eqref{equ:system_far_field_with_damping}, we obtain the modified continuity equation~\eqref{equ:modified-continuity} introduced in section~\ref{section:absorbing-layer}.

\begin{figure}%
\centering%
    \includegraphics[width=6cm]{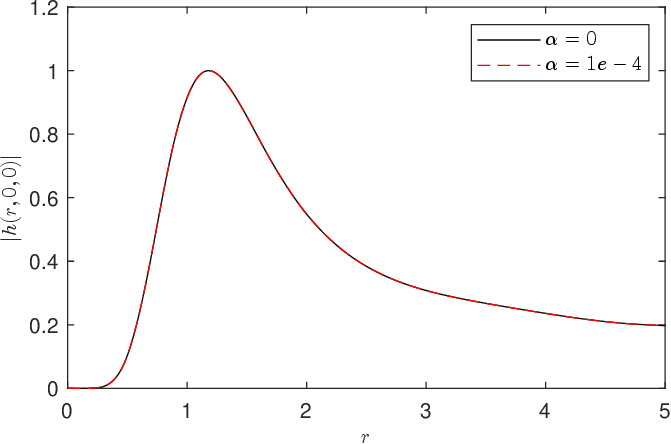}%
    \includegraphics[width=6cm]{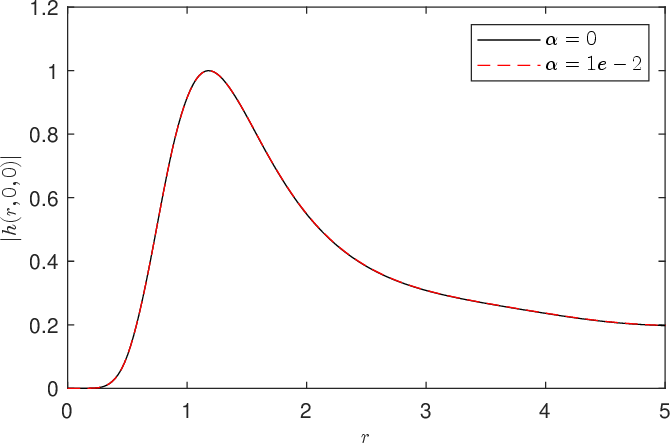}%
\par%
    \includegraphics[width=6cm]{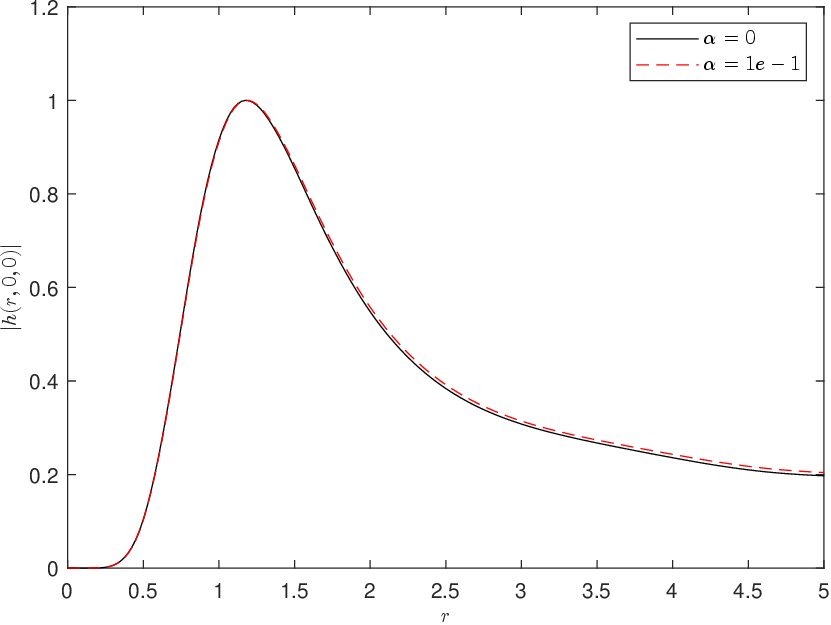}%
    \includegraphics[width=6cm]{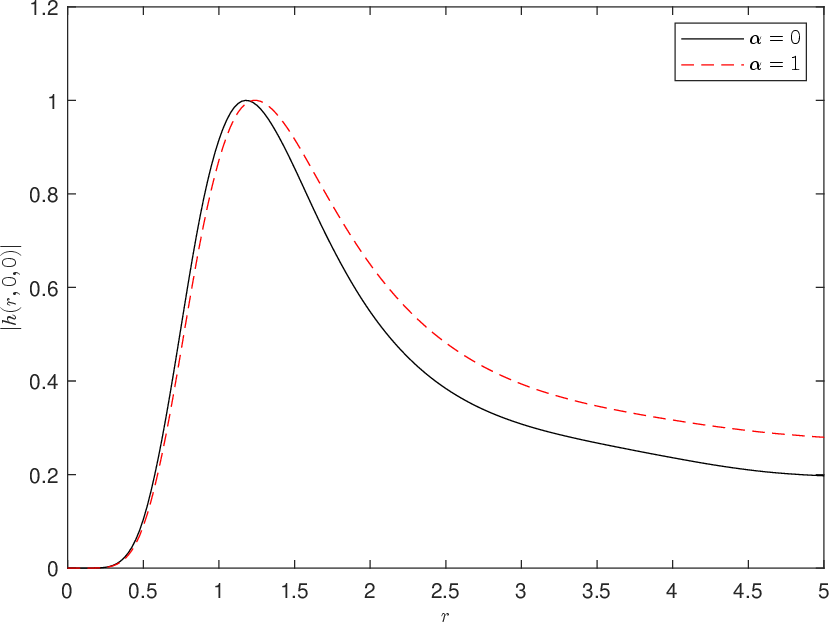}%
\caption{Comparison of the structure of the eigenmode $|h(r, 0, 0)|$ for $m = 7$, $F = 0.3$ and different value of the compressibility $\alpha = 1/c^2$. }%
\label{fig:comparison_mode_different_alpha}%
\end{figure}

Figure~\ref{fig:comparison_mode_different_alpha}
and table~\ref{table:comp_evals_with_alpha}
\begin{table}
\centering
 \begin{tabular}{c c c c c c} 
 \hline
 $\alpha$ & $0$ & $10^{-4}$ & $10^{-2}$ & $10^{-1}$ & $1$ \\ [0.5ex]
 \hline\hline
 $\Real(\omega)$ & -1.3769 & -1.3769 & -1.3759 & -1.3662 & -1.2711 \\
 $\Imag(\omega)$ & -0.0138 & -0.0138 & -0.0139 & -0.0147 & -0.0241 \\
 \hline
 \end{tabular}
 \caption{Eigenvalues as function of the compressibility $\alpha = \frac{1}{c^2}$ for a radiating mode, for $m = 7$, and $F = 0.3$.}
 \label{table:comp_evals_with_alpha}
\end{table}%
show the effect of the compressibility $\alpha = 1/c^2$ on the structure of the radiating eigenmode and the eigenvalue respectively, for parameters $m = 7$ and $F = 0.3$.  From both, it is clear that the eigenmode computed does not vary significantly as $\alpha \to 0$, and so we conclude that the damping properties of the $\xi(r)p$ term are carried over in the limit $\alpha \to 0$.
%
%
% COMPARISON MODES NEWTON'S BUCKET PROBLEM AND CODE VALID.
%
%
\section{Code validation and comparison with results for other background flow}
\label{appendix:code_validation_SB_rotation}

Here we provide evidence of the correctness of our numerical methods. As a first test, we have validated our code against the results of Mougel, {\em et al.} for surface waves on a Rankine vortex in a bounded domain \cite[Fig.~5]{mougel-2014}. We have adapted our code to this base flow and geometry by including a rigid wall at finite radius, and therefore not using a damping region to model an unbounded domain. With these changes, we find excellent agreement between our computations and theirs. For example, at $Fr = 1.502$ they find an unstable surface wave with eigenvalue $2.6101 + 0.0025313i$ (J. Mougel, private communication). We confirm this instability and find $\omega = 2.6111+0.004556i$. The moduli of these eigenvalues agree to better than 0.04\%. Mougel, {\em et al.} include a small amount of viscous damping in their computations, while we do not, and this can account for the small absolute difference in the imaginary parts of the respective eigenvalues. 
By including an artificial viscous term (with artificial viscosity $\nu$) of the form $\nu \partial^{2}_{x} u$ in the right-hand side of equation \eqref{equ:x_momentum_xy_vars}, we have confirmed the sensitivity of the small imaginary part of the eigenvalue to damping, but we have not attempted to reproduce the exact damping used by Mougel, {\em et al.}.

\begin{figure}%
\centering%
    \includegraphics[width=4.2cm]{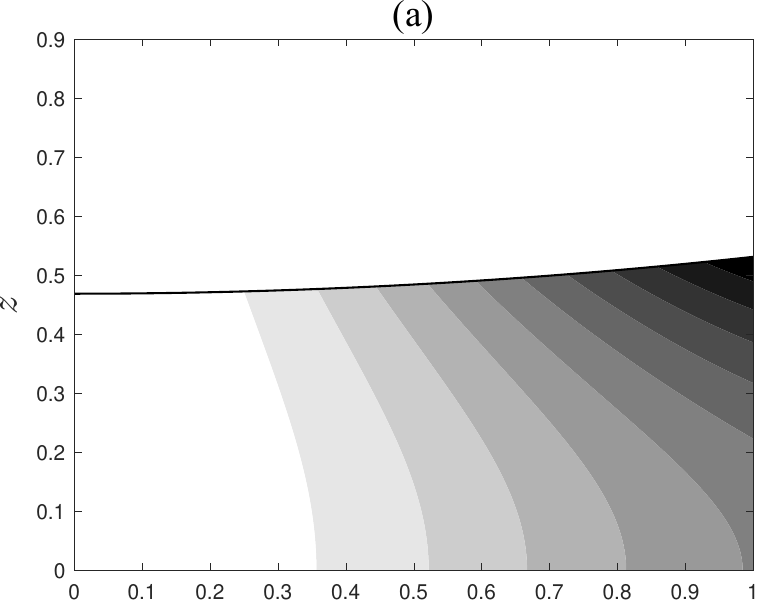}% 
    \hspace{0.5cm}
    \includegraphics[width=4cm]{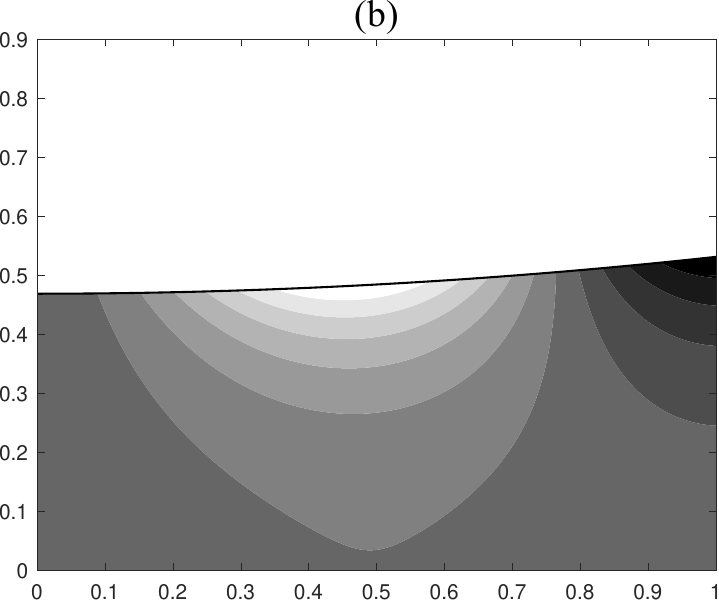}%
    \hspace{0.5cm}
    \includegraphics[width=4cm]{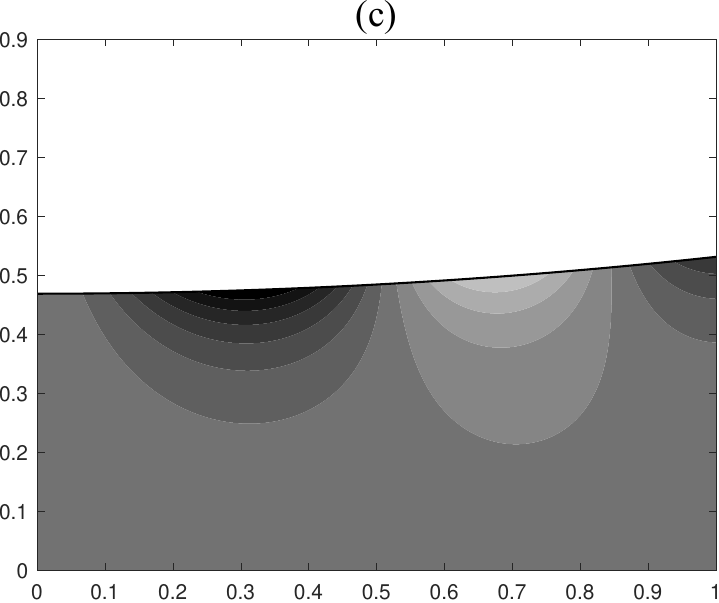}%
\par%
\vspace{0.5cm}
    \includegraphics[width=4.2cm]{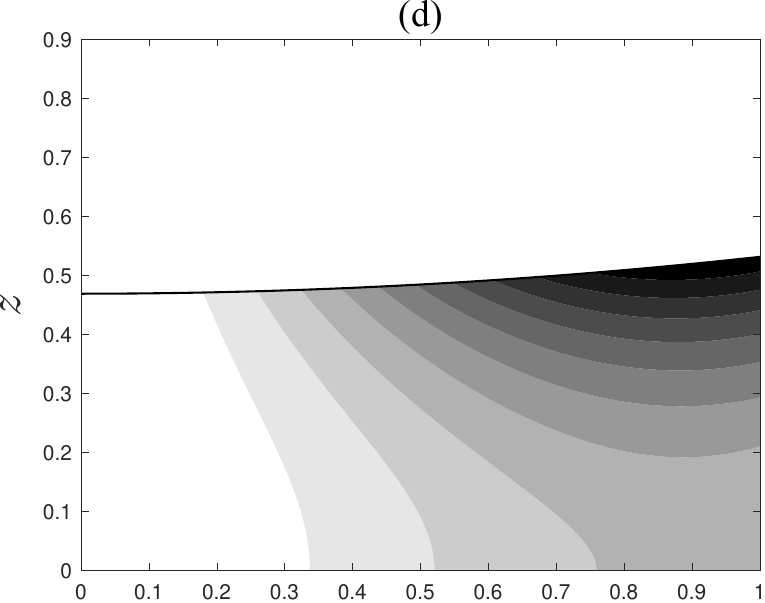}% 
    \hspace{0.5cm}
    \includegraphics[width=4cm]{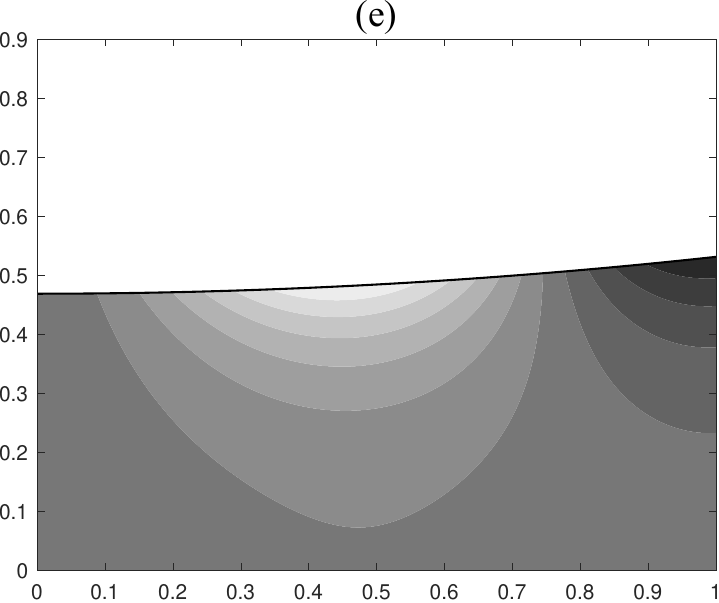}%
    \hspace{0.5cm}
    \includegraphics[width=4cm]{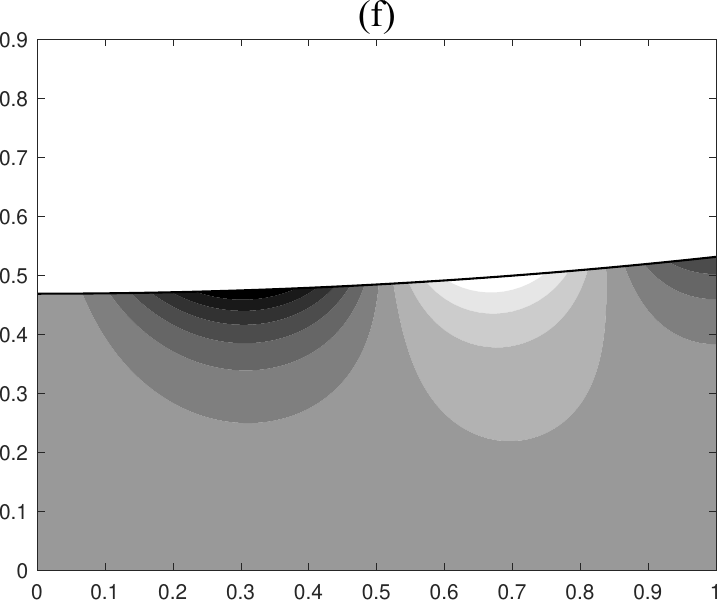}%
\par%
\vspace{0.5cm}
    \includegraphics[width=4.2cm]{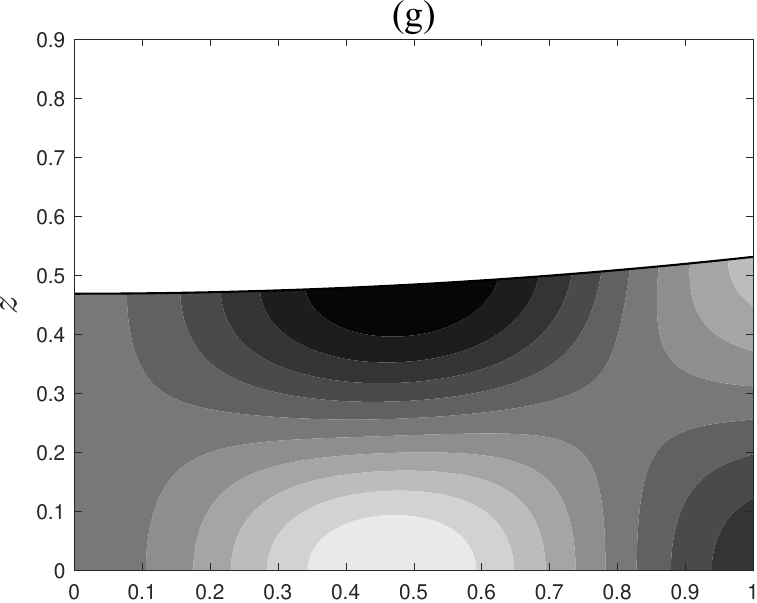}% 
    \hspace{0.5cm}
    \includegraphics[width=4cm]{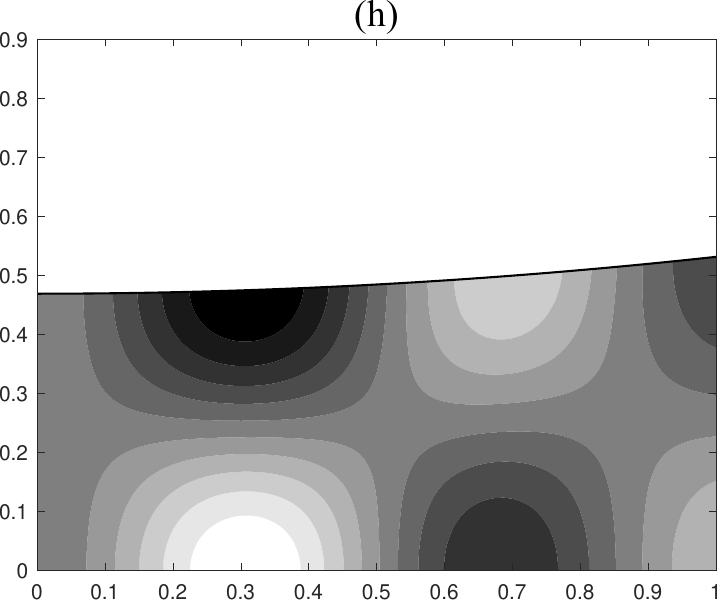}%
    \hspace{0.5cm}
    \includegraphics[width=4cm]{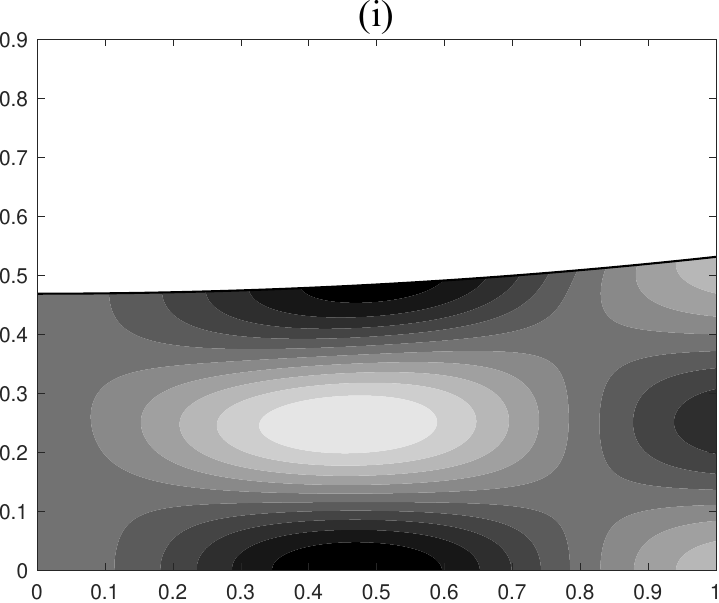}%
\par%
\vspace{0.5cm}
    \includegraphics[width=4.2cm]{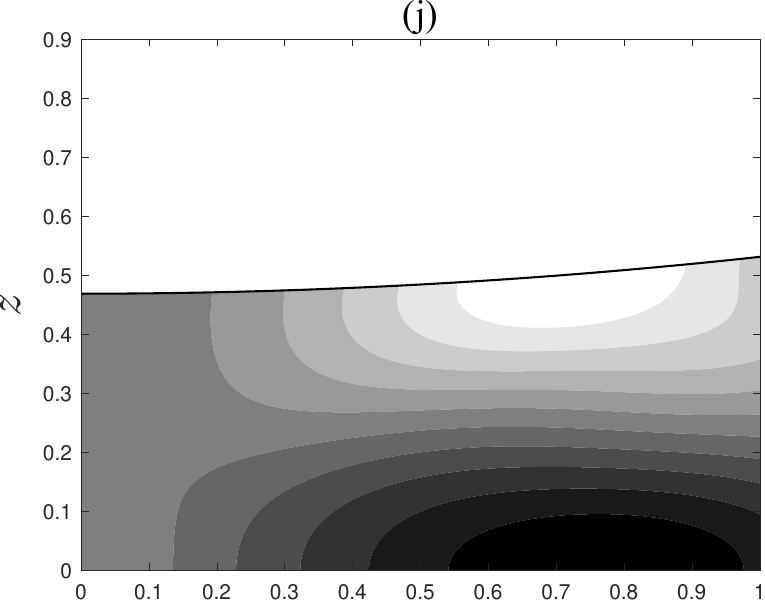}%
    \hspace{0.5cm}
    \includegraphics[width=4cm]{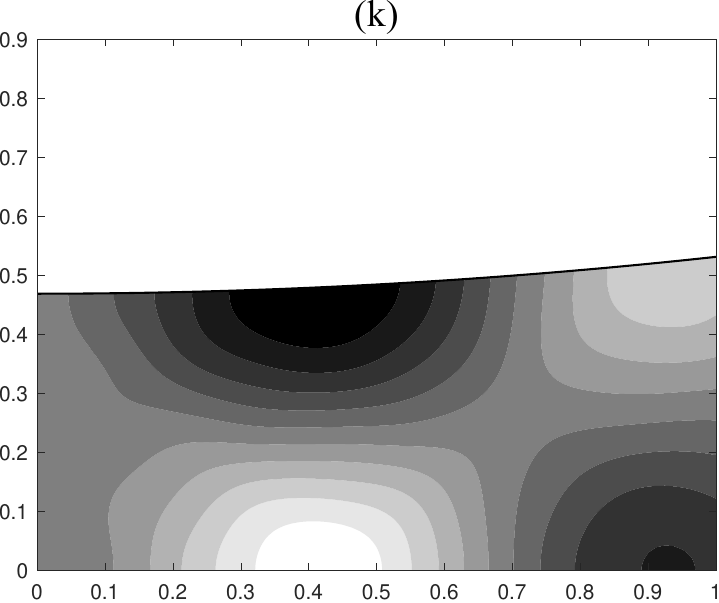}%
    \hspace{0.5cm}
    \includegraphics[width=4cm]{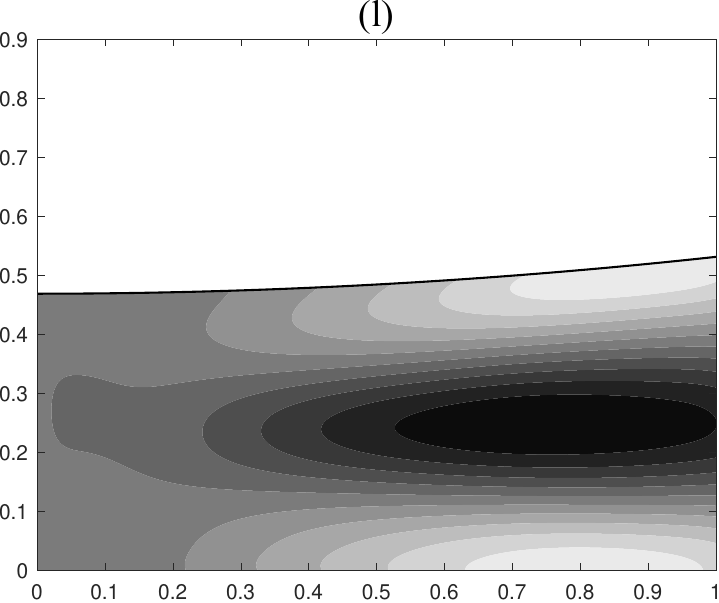}%
\par%
\vspace{0.5cm}
    \includegraphics[width=4.2cm]{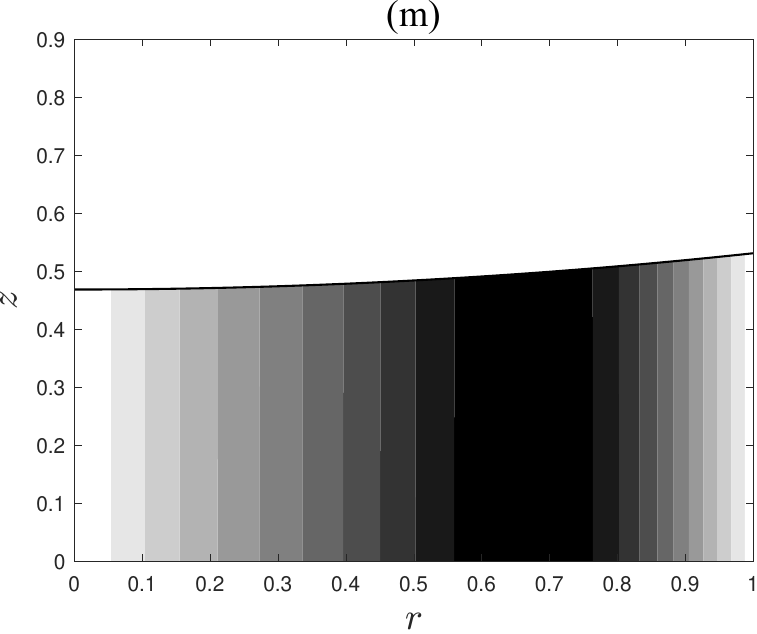}% 
    \hspace{0.5cm}
    \includegraphics[width=4cm]{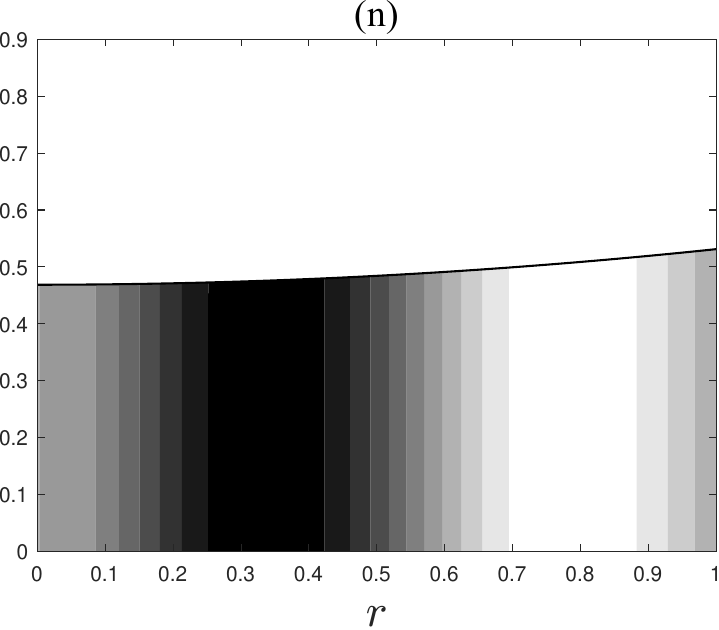}%
    \hspace{0.5cm}
    \includegraphics[width=4cm]{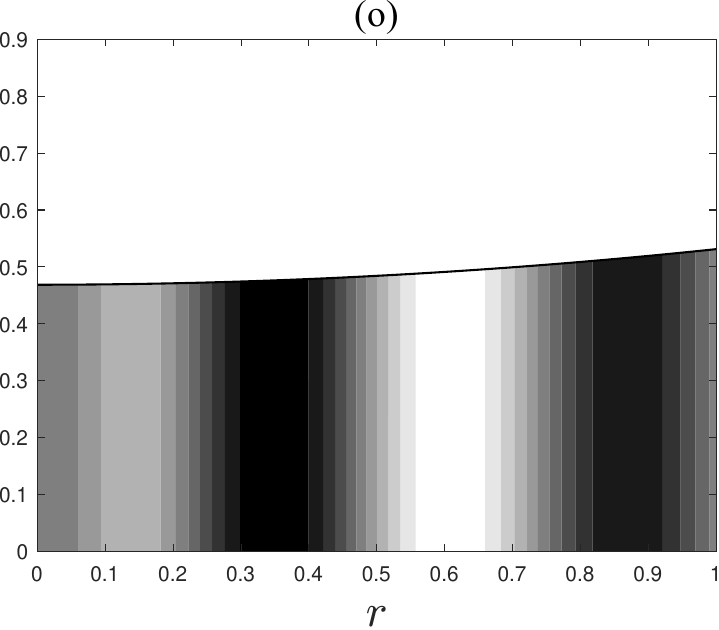}%
\caption{Eigenmodes comparison with Newton's Bucket Problem \citet{mougel-2015} using their same parameters and our artificial viscosity $\nu = 0.01$. 
(a): $\lambda = 4.2897 - 0.008\I$. (b): $\lambda = 7.4359 - 0.0123\I$. (c): $\lambda = 9.0436 - 0.0368\I$. (d): $\lambda = -5.2014 - 0.0029\I$. (e): $\lambda = -7.4580 - 0.0156\I$. (f): $\lambda = -9.0349 - 0.0408\I$. 
(g): $\lambda = 1.4077 - 0.0112\I$. (h): $\lambda = 1.0974 - 0.0254\I$. (i): $\lambda = 1.7833 - 0.0180\I$. (j): $\lambda = -1.6985 - 0.0111\I$. (k): $\lambda = -1.3292 - 0.0304\I$. (l): $\lambda = -1.9142 - 0.0122\I$. (m): $\lambda = -0.0346 - 0.0086\I$. (n): $\lambda = -0.0124 - 0.0139\I$. (o): $\lambda = -0.0077 - 0.0166\I$.
First two rows represent surface gravity modes. Third and fourth row inertial modes. Last row Rossby modes.}
\label{fig:comparison_modes_newtons_bucket}%
\end{figure}

As a second test case, we have validated our code against results of \citet{mougel-2015} for Newton's bucket problem: solid body rotation with a rigid walls at $r=R$. We have adapted our code to this case. Results from our numerics are plotted in figure~\ref{fig:comparison_modes_newtons_bucket} and can be compared with Fig.~3 of \citet[][p.~223]{mougel-2015}. For this case, we adopt the same non-dimensionalization as \citet{mougel-2015}, so that we can directly compare our numerical solutions to theirs.

To obtain smooth and well-resolved eigenfunctions for Newton's bucket, we have again added the artificial viscous term explained just above for the Rankine vortex. Excellent agreement is found in this case as well, giving further confidence in the correctness of our numerics. For example, by setting our artificial viscosity $\nu = 0.01$, the eigenvalue of the first gravity mode from our computation is $\lambda = 4.2897 - 0.008\I$, whereas they found $\lambda = 4.290 - 0.0014 \I$. For the first inertial mode we find $\lambda = 1.4077 - 0.0112\I$, against their result $\lambda = 1.408 - 0.0083\I$. Finally for the first Rossby mode we have $\lambda = -0.0346 - 0.0086\I$, whereas they found $\lambda = -0.0372 - 0.0021\I$.

As a third and final test case, we
have computed the stability the steady Lamb-Oseen vortex for the shallow-water equations presented in \citet{ford}. Using the same convention as Ford, we obtain an unstable linear mode at the following set of parameters: $F = 1$, $m = 2$, and $f = 0$, where $f$ is the Coriolis parameter. Specifically, for these parameter values our code returns the most unstable eigenvalue $\omega = 0.2606 + 0.0017\I$, and this very agrees well with the eigenvalues plotted in \citet[][figure 4]{ford}, even though the vorticity profile used here and by Ford are not the identical. We again confirm that our code finds instability where it is known to exist.

\section{Results without the base free surface deformation}
\label{appendix:results_without_free_surface_deformation}

\begin{figure}%
\centering%
    \includegraphics[]{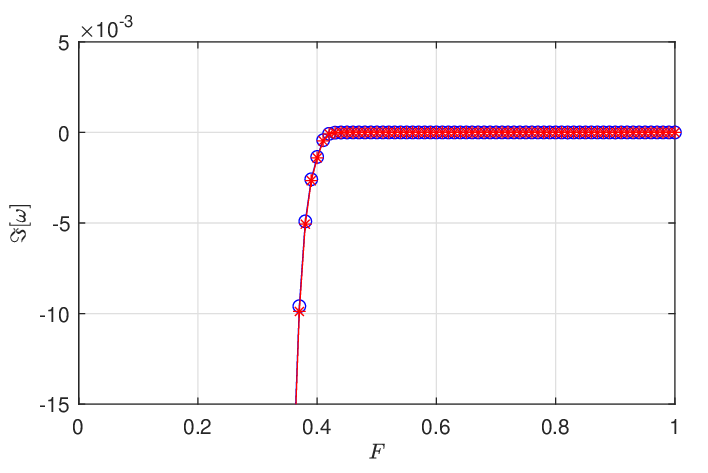}%
    \par
    \includegraphics[]{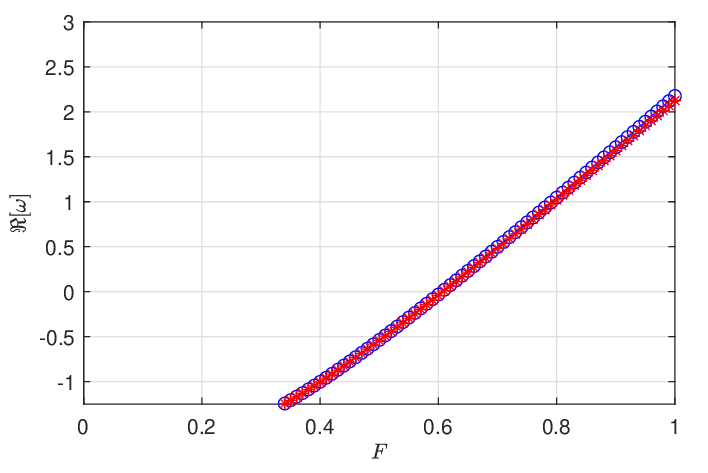}%
    \caption{Imaginary and real part of the eigenvalues as function of the Froude number for modes with $m = 7$ and $n = 1$. Blue-circle curve: full numerics; red-asterisk curve: without the free surface contribution.}%
\label{fig:spectrum_curves_with_and_without_FS}%
\end{figure}

\begin{figure}%
\centering%
    \includegraphics[]{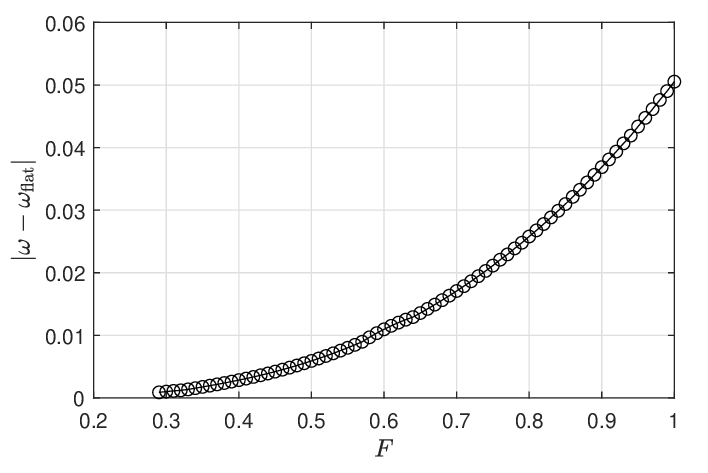}%
    \caption{Absolute difference between eigenvalues computed with and without the free surface contribution, $|\omega - \omega_{\mathrm{flat}}|$, for the first radial mode $n=1$ and $m = 7$. }%
\label{fig:error_evals_with_and_without_FS}%
\end{figure}

In this appendix, we demonstrate that the dominant contribution to the trapping of modes by the vortex shown in section~\S\ref{sec:results} is the swirl of the base flow, and that the deformation of the base free surface
has little effect itself. In the following we artificially assume the base free surface to be flat, meaning that $h_{0}(r) = h_{\infty}$ and $h'_{0}(r) = 0$. All the remaining terms in the governing equations and boundary conditions do not change.  Solving the eigenvalue problem in this case, we compare the eigenvalues to the complete case where the base free surface deformation has been taken into account in figure~\ref{fig:spectrum_curves_with_and_without_FS}.
Little difference is seen between the two.  Indeed, figure~\ref{fig:error_evals_with_and_without_FS}
plots the difference in eigenvalues as function of the Froude number; i.e.\ $|\omega - \omega_{\mathrm{flat}}|$, where $\omega_{\mathrm{flat}}$ are the eigenvalues without the free surface variation. It is clear that for the range of Froude numbers considered, the two agree remarkably closely.

\begin{figure}%
\centering%
    \includegraphics[]{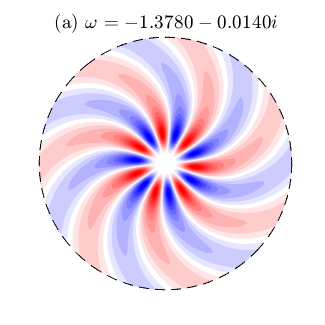}%
    \includegraphics[]{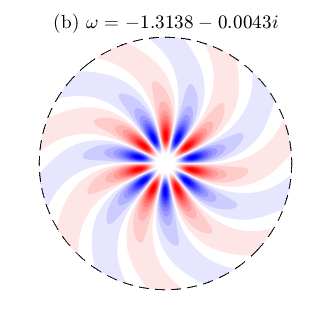}%
\par%
    \includegraphics[]{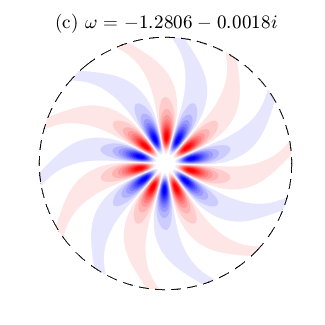}%
    \includegraphics[]{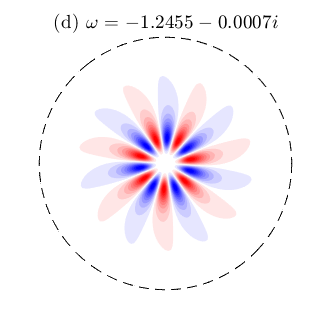}%
\par%
    \includegraphics[]{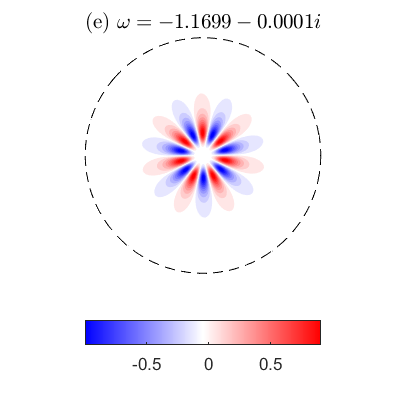}%
\caption{Plots of the perturbation free surface height $h(r, \theta, t=0) = \Real[\phi(r, h_{0}(r))\exp\{\I m\theta\}]$ for $m = 7$ without the base free surface deformation, to be compared against figure~\ref{fig:free_surface_above_m7_varying_F}. 
(a) $F = 0.3$. (b) $F = 0.32$. (c) $F = 0.33$. (d) $F = 0.34$. (e) $F = 0.36$.
All modes displayed rotate clockwise, i.e.\ against the vortex flow.}%
\label{fig:free_surface_above_m7_varying_F_without_FS}%
\end{figure}

\begin{figure}%
\centering%
    \includegraphics[]{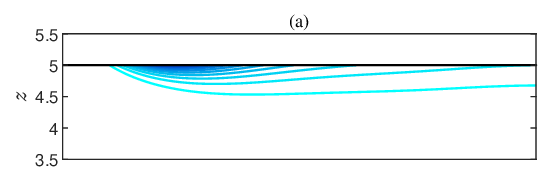}%
\par%
    \includegraphics[]{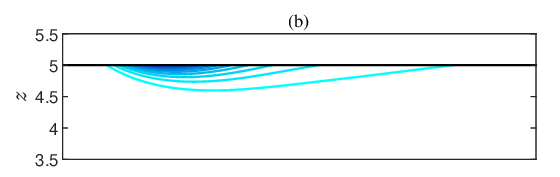}%
\par%
    \includegraphics[]{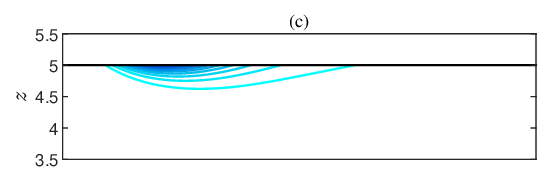}%
\par%
    \includegraphics[]{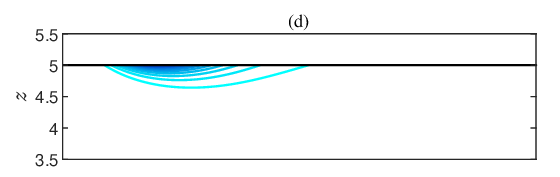}%
\par%
    \includegraphics[]{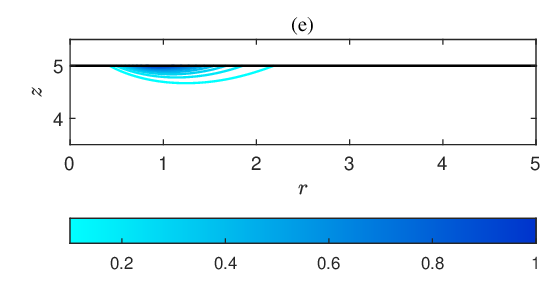}%
\caption{Plots of the pressure distribution $|\phi_{m}(r, z)|$ for $m = 7$ without the base free surface deformation, to be compared against figure~\ref{fig:pressure_contour_radiating_regime}. (a) $F = 0.3$. (b) $F = 0.32$. (c) $F = 0.33$. (d) $F = 0.34$. (e) $F = 0.36$.}%
\label{fig:pressure_contour_radiating_regime_without_FS}%
\end{figure}

In terms of the eigenfunctions, the free surface perturbation is displayed in figure~\ref{fig:free_surface_above_m7_varying_F_without_FS},
which should be compared against the free surface perturbation with a base free surface height variation plotted in figure~\ref{fig:free_surface_above_m7_varying_F}; the two figures can be seen to be almost identical. The corresponding structure in the $r-z$ plane is shown in figure~\ref{fig:pressure_contour_radiating_regime_without_FS},
which should be compared against figure~\ref{fig:pressure_contour_radiating_regime} in the results section; again, the two figures are almost identical apart from the base flow height.

We therefore conclude that the driving mechanism behind the trapping of modes by the vortex is the rotation of the base flow, and not the free surface deformation.

\bibliography{bibliography}

\end{document}